\documentclass[twocolumn,nofootinbib,pra,preprintnumbers]{revtex4}%
\usepackage{graphicx} %
\usepackage{subfigure}
\usepackage{hyperref}%

\usepackage{bm}
\usepackage{amsmath}
\usepackage{amssymb}
\usepackage{amsfonts}
\usepackage{fancyhdr}
\usepackage{appendix}
\usepackage{makeidx}

\begin{document}
\preprint{ITP-UU-07/66}
\preprint{SPIN-07/52}

\title{Hermitian Gravity and Cosmology}

\author{Christiaan Mantz}
\email{CLMMantz@hotmail.com}
\author{Tomislav Prokopec}
\email{T.Prokopec@phys.uu.nl}
\affiliation{Institute for
Theoretical Physics, University of Utrecht, Princetonplein 5, P.O.
Box 80.006, 3508 TA Utrecht, The Netherlands}
\date{\textit{\today}}

\begin{abstract}
\noindent In an attempt to generalize general relativity, we
propose a new Hermitian theory of gravity. Space-time is
generalized to space-time-momentum-energy and both the principles
of general covariance and equivalence are extended. The theory is
endowed with a Hermitian metric on a complex manifold. The
Hermitian metric contains, apart from the symmetric metric, an
anti-symmetric part, which describes dynamical torsion. The
causality structure is changed in a way such that there is a
minimal time for events to be in causal contact and a maximal
radius for a non-local instantaneous causally related volume. The
speed of light can exceed the conventional speed of light in
non-inertial frames and accelerations are bounded. We have
indications that the theory of Hermitian gravity yields general
relativity at large scales and a theory equivalent to general
relativity at very small scales, where the momenta and energies
are very large. As an example, we study cosmology in Hermitian
gravity, where  matter is described by two scalar fields. While at
late times Hermitian gravity reproduces the standard cosmological
FLRW models, at early times it differs significantly: quite
generically the Universe of Hermitian cosmology exhibits a bounce
where a maximal expansion rate (Ricci curvature) is attained.
Moreover, we prove that no cosmological constant is permitted at
the classical level within our model of Hermitian cosmology.
\end{abstract}
%
\maketitle %

%
%
%
%
\pagenumbering{arabic}
\section{Motivation}

\noindent According to Albert Einstein's principle of relativity
the laws of physics are independent of system of reference. The
principle of equivalence states that an observer cannot tell
whether he is accelerating or placed in a gravitational field. If
we follow Einstein's principle of relativity closely one could
argue that there must be a similar principle of equivalence
between rotating observers and observers placed in a torsion
field. A torsion field is a gravitational field which causes
observers to rotate~\cite{Mao:2006bb}. Theories of generalized
gravity, in which dynamical torsion is present, have been
proposed, using the standard principle of covariance. The theory
proposed by Moffat~\cite{Moffat:1978tr} has unsatisfactory
properties~\cite{Janssen:2006jx}\cite{Damour:1991ru}, which caused
many to give up dynamical torsion.

In this article we propose a theory of dynamical torsion, by not
only extending the principle of equivalence, but also by extending
the principle of covariance. We generalize space-time to
space-time-momentum-energy, imposing the reciprocity symmetry as a
symmetry between space-time and momentum-energy, as was suggested
by Max Born in 1938~\cite{Born1938Reciprocity}. Max Born's
original motivation behind the idea that the laws of physics
should be invariant under the reciprocity transformation was that
position and momentum operators of quantum mechanics obey the
reciprocity symmetry transformation. Hence a theory unifying
quantum mechanics and gravity should also be invariant under the
reciprocity transformation. Our aim is to formulate this new
theory, incorporating the reciprocity symmetry and simultaneously
satisfying an extended principle of general covariance. We will
leave the quantum aspect of the theory for future work.

\subsection{The Reciprocity Principle}
\noindent According to Max Born the laws of physics are invariant
under the reciprocity transformation, which is given by
\begin{eqnarray}\label{eq:reciprocity transformation}
    x^\mu \rightarrow p^\mu
    \phantom{halloda}
    p^\mu \rightarrow - x^\mu,
\end{eqnarray}
where $x^\mu$ and $p^\mu$ are the four vectors ($ct,\vec{x}$) and
($\frac{E}{c},\vec{p}$), respectively. The components of the
angular momentum
\begin{eqnarray}\nonumber
    x_\mu p_\nu - p_\mu x_\nu
    =
    M_{\mu\nu}
\end{eqnarray}
are indeed invariant under the reciprocity transformation. Since
torsion couples to angular momentum \cite{Mao:2006bb}, it seems
natural to demand that a new theory describing dynamical torsion
should be invariant under the reciprocity transformation.

Note that, when quantizing this new theory, which
we leave for future work, the commutation relations from
quantum mechanics
\begin{eqnarray}
    {\hat{x}}^\mu {\hat{p}}_\nu - {\hat{p}}_\nu {\hat{x}}^\mu
    =
    i\hbar \delta^\mu_\nu
\label{eq:commutation relations}
\end{eqnarray}
are also invariant under the reciprocity relation.

\subsection{General Relativity and the Reciprocity Principle}
\noindent The theory of general relativity describes our universe
at large scales (at the moment we are not considering
 cosmological issues such as dark matter and dark energy)
and it generalizes classical mechanical ideas as orbits,
instead of wave functions, in order to describe particles.
The four dimensional line element
\begin{eqnarray}\label{eq:line element spacetime}
    d s^2
    =
    g_{\mu\nu}d x^\mu d x^\nu
\end{eqnarray}
is a fundamental notion in the theory of general relativity. It is
clear that general relativity and the way distances are determined
(\ref{eq:line element spacetime}) breaks the reciprocity symmetry
(\ref{eq:reciprocity transformation}). Demanding that the theory,
unifying quantum  mechanics and general relativity, should respect
the principle of reciprocity, we can state a four dimensional
momentum-energy line element
\begin{eqnarray}\label{eq:line element MomentumEnergy}
    d \sigma^2
    =
    \gamma^{\mu\nu}d p_\mu d p_\nu,
\end{eqnarray}
which should dominate over the space-time line element
(\ref{eq:line element spacetime}), whenever the momenta are very
large compared to this position length scale.

According to the classical laws the momentum $p^\mu$ is given by
$m \dot{x^\mu}$, which corresponds to the tangent vector of the
path taken. The idea of having a tangent space at each point of
the manifold, corresponding to the physical idea of the momentum
as tangent vector, is clearly only applicable in the classical
realm of physics, when momenta are small compared to distances.
For the sake of brevity, Max Born called this scale, at which the
theory of general relativity is valid the molar
world~\cite{Born1938Reciprocity}, while he called the "small world",
which is described by the momentum energy line element
(\ref{eq:line element MomentumEnergy}) the nuclear world. The
world, which lies in between these worlds on the energy-momentum
and space-time scales, is familiarly called the quantum world,
which is the realm of quantum gravity.

Since general relativity is governed by Einstein's equations
\begin{eqnarray}\nonumber
   R_{\mu\nu} - \frac{1}{2}g_{\mu\nu}R - \Lambda g_{\mu\nu}
   =
   \kappa T_{\mu\nu},
\end{eqnarray}
we can state via the principle of reciprocity the reciprocal
Einstein equations
\begin{eqnarray}\nonumber
   P^{\mu\nu} - \frac{1}{2}\gamma^{\mu\nu}P - \Lambda'\gamma^{\mu\nu}
   =
   \kappa'{T'}^{\mu\nu},
\end{eqnarray}
which are supposed to govern the momentum-energy curvature of the
nuclear world.

We now have a vague idea of how the theory should behave in
certain limits and the principles it should obey, namely the
principle of reciprocity and the generalized principle of general
covariance. Our goal is now to construct a theory which obeys all
these limits and principles. The principle of general covariance
suggests that there should exist a space-time-momentum-energy line
element that specifies a corresponding space-time-momentum-energy
interval, which is absolute in the sense that all observers would
agree on it; interpreting the momentum-energy coordinates as
coordinates, specifying non inertial frames, all relatively non
inertial moving observers should agree on the measured
space-time-momentum-energy interval. While our proposal extends
the standard covariance principle, proposed by Einstein in 1905
\cite{Einstein1905SR}, it necessarily breaks this principle of
covariance by introducing energy-momentum into the distance
measurements. Space-time then becomes a relative space with
respect to observers moving non inertially with respect to each
other and becomes absolute only in the limit of relatively
inertial moving observers.

\section{Almost Complex Structure}

\noindent Clearly we need a metric on a manifold to mathematically
describe space-time-momentum-energy curvature. In order to build a
theory which is reciprocal in momentum and space and at the same
time reduces to the theory of general relativity in the molar
limit, we need complex manifolds with a Hermitian metric. This is
the case because the Hermitian metric is defined such that it is
invariant under the reciprocity transformation
(\ref{eq:reciprocity transformation}). Any $2d$ dimensional
manifold, with a $d$ $x$ and $d$ $y$ coordinates, locally admits a
tensor field $J$~\cite{Nakahara:2003nw}, which maps the tangent space of
the manifold into itself, $J_p : T_p M \rightarrow  T_p M $, in
the following manner
\begin{eqnarray}\label{eq:reciprocity transformation by J operation}
    J\left(x^\mu\right) \rightarrow y^\mu
    \phantom{hallo}
    J\left(y^\mu\right) \rightarrow - x^\mu,
\end{eqnarray}
where the index $\mu$ runs from $0$ to $d-1$. It is clear that
this map is equivalent to the reciprocity transformation
(\ref{eq:reciprocity transformation}), if the $y$ coordinate is
interpreted as the energy-momentum coordinate. The map $J$, also
known as the "almost complex structure" operator, may be defined
globally on a complex manifold and then it specifies completely
the complex structure of the manifold. A metric C, which is
invariant under the action of this $J$ operator in the following
way
\begin{eqnarray}\nonumber
    C_p \left( J_p Z ,  J_p W \right)
    =
     C_p \left( Z ,  W \right)
\,,
\end{eqnarray}
is a Hermitian metric, where $Z,W \in  T_p M$ and $T_p M$ is the
complexified tangent space~\cite{Nakahara:2003nw}. The action of the
almost complex structure operator on the basis vectors of the
complexified tangent space follows from the definitions of the
almost complex structure map and these basis vectors
\begin{eqnarray}\label{eq:almost complex structure z zbar operation}
    J_p\left(\frac{\partial}{\partial z^\mu}\right)
    =
    i \frac{\partial}{\partial z^\mu}
    \phantom{halloda}
    J_p\left(\frac{\partial}{\partial \bar{z}^\mu}\right)
    =
    - i \frac{\partial}{\partial \bar{z}^\mu}.
\end{eqnarray}
Consider a full complex metric
\begin{eqnarray}\label{complex metric}
   C
   =
   C_{\mu\nu} d z^{\mu} \otimes d z^{\nu}
   &+&
   C_{\mu\bar\nu} d z^{\mu} \otimes d z^{\bar\nu}
\\\nonumber
   +
   C_{\bar\mu\nu} d z^{\bar\mu} \otimes d z^{\nu}
   &+&
   C_{\bar\mu\bar\nu} d z^{\bar\mu} \otimes d z^{\bar\nu}
\,.
\end{eqnarray}
A Hermitian metric is a complex metric which has
-- as a consequence of the reciprocity symmetry --
vanishing $C_{\mu\nu}$ and $C_{\bar\mu\bar\nu}$ components:
\begin{eqnarray}\nonumber
   C
   =
   C_{\mu\bar\nu} d z^{\mu} \otimes d z^{\bar\nu}
   +
   C_{\bar{\mu}\nu} d z^{\bar{\mu}} \otimes d z^{\nu},
\end{eqnarray}
where barred indices $z^{\bar{\mu}} \equiv {\bar{z}}^{\mu}$ denote complex conjugation.
The basic definitions of complex manifolds do not differ too much
from the usual definitions of a manifold, except for the fact that
the complex manifold is locally homeomorphic to the complex space
$\mathbb{C}^m$ and the coordinate transformations are holomorphic
and hence satisfy the Cauchy-Riemann equations.

\section{The Hermitian Metric}
\noindent We can write the Hermitian line element in eight
dimensional form
\begin{eqnarray}\nonumber
    ds^2 &=&  d\boldsymbol{z}^T\cdot \boldsymbol{C} \cdot d\boldsymbol{z}
    =
   d\boldsymbol{z}^m\boldsymbol{C}_{mn}d\boldsymbol{z}^n
   \\\label{eq:line element hermitian metric subsection}
   &=&
        (dz^{\mu},
        dz^{\bar{\mu}})
    \left(\begin{array}{cc}
       0 & C_{\mu\bar{\nu}}\\
        C_{\bar{\mu}\nu} & 0
    \end{array}\right)
    \left(\begin{array}{c}
        dz^{\nu}\\
        dz^{\bar{\nu}}
    \end{array}\right),
\end{eqnarray}
where the Latin indices can take the values
$0,1,...,$$d-1$,$\bar{0},\bar{1},...,\overline{d-1}$, where the
Greek indices can take the values $0,1,...,d-1$ and where the
number $d$ is the complex dimension of the complex manifold. The
entries of the metric $C_{mn}$ are functions of holomorphic and
anti antiholomorphic vielbeins defined as follows~\footnote{Here the
Latin indices $a,b$ run from $0,1,...,d-1$, since they represent
local indices, and $\eta = $ diag$(-1,1,1,1)$. We
will the discuss the meaning of these indices later. 
}
\begin{align}\label{eq:vielbein complex metric in terms of vielbeins}
    &C_{\mu\bar{\nu}}
    =
    e(z)_{\mu}^{\phantom{\mu}a}\eta_{ab}e(\bar{z})_{\bar{\nu}}^{\phantom{\bar{\nu}}b}\\\nonumber
    &C_{\bar{\mu}\nu}
    =
    e(\bar{z})_{\bar{\mu}}^{\phantom{\bar{\nu}}a} \eta_{a b}e(z)_{\nu}^{\phantom{\nu}
    b},
\end{align}
where $\eta_{ab}  \equiv $ diag$(-1,1,1,1)$.
For completeness, we also quote the other two elements of
the full complex metric~(\ref{complex metric}),
\begin{eqnarray}\label{eq:complex metric in terms of vielbeins:2}
    C_{\mu{\nu}}
    &=&
    e(z)_{\mu}^{\phantom{\mu}a}\eta_{ab}e({z})_{{\nu}}^{\phantom{\bar{\nu}}b}
\\\nonumber
    C_{\bar{\mu}\bar\nu}
    &=&
    e(\bar{z})_{\bar{\mu}}^{\phantom{\bar{\nu}}a}
          \eta_{a b}e(\bar z)_{\bar\nu}^{\phantom{\nu}
    b}
\,.
\end{eqnarray}
Note that holomorphy of the vielbeins $e_\mu^{\,a}=e_\mu^{\,a}(z)$
implies the reality condition for the line element,
\begin{eqnarray}\label{eq:reality condition of the line element}
    {ds^2}^\dag = ds^2,
\end{eqnarray}
since it implies the following definition for complex conjugation
of the metric,
\begin{eqnarray}\nonumber
    {C_{\bar{\mu}\nu}}^* \equiv \overline{ C_{\bar{\mu}\nu}} =
    C_{\mu\bar{\nu}}.
\end{eqnarray}
We can also write the Hermitian line element in its familiar four
dimensional form
\begin{eqnarray}\label{eq:Hermitian familiar form line element}
    ds^2
    =
    2d z^T \cdot C \cdot d \bar{z}
    =
    2dz^\mu C_{\mu\bar{\nu}}d z^{\bar{\nu}},
\end{eqnarray}
which is a logical extension of the complex inner product $\langle
w,v\rangle = \bar{w_i}v_i $. Note that, in this familiar form
(\ref{eq:Hermitian familiar form line element}), the Hermitian
metric is actually Hermitian in the usual way, $C= C^\dagger $,
since the line element is real. We shall see that the eight
dimensional notation (\ref{eq:line element hermitian metric
subsection}), through which the Hermitian metric takes its
symmetric form, $ \bm C^T = \bm C $, is very handy for obtaining
the equations of motion for this theory. We can define the $z^\mu$
and ${\bar{z}}^{\bar{\mu}}$ coordinates in terms of $x^\mu$ and
$y^{\check{\mu}}$, as follows\footnote{Checks ( $\check{}$ ) are
put on indices to denote the imaginary part of a coordinate or on
indices of objects, which are projected onto its basis vector.}
\begin{eqnarray}\nonumber
    z^\mu
    =
    \frac{1}{\sqrt{2}}(x^\mu+ i y^{\check{\mu}})
    \phantom{hallo}
    \frac{\partial}{\partial z^\mu}
    &=&
    \frac{1}{\sqrt{2}}\Big(\frac{\partial}{\partial x^\mu}
       - i \frac{\partial}{\partial y^{\check{\mu}}}\Big)
\end{eqnarray}
and their complex conjugates. This implies the following
decomposition of complex vielbeins in their real, ${e_R}^a_\mu$,
and imaginary, ${e_I}^a_{\check{\mu}}$, parts in the following
manner
\begin{eqnarray}\nonumber
    e_a^\mu
    &=&
    {e_R}_a^\mu + i {e_I}_a^{\check{\mu}}\,,
    \phantom{hallo}
    \overline{e_a^{\mu}}
    =
    e_a^{\bar{\mu}} = {e_R}_a^\mu - i{e_I}_a^{\check{\mu}}
    \\\label{eq:viebein decompositions}
    e^a_\mu
    &=&
    {e_R}^a_\mu - i {e_I}^a_{\check{\mu}}\,,
    \phantom{hallo}
    \overline{e^a_{\mu}}
    =
    e^a_{\bar{\mu}} = {e_R}^a_\mu + i{e_I}^a_{\check{\mu}}.
\end{eqnarray}
Vielbeins are holomorphic functions, and thus transform as
holomorphic vectors (we consider only the transformation of the
Greek indices for this purpose),
\begin{eqnarray}\label{tetrad:coord.transformations}
  &&e_\mu^a(z^\nu) \rightarrow
    \tilde e_\mu^a(w^\nu) = \frac{\partial z^\alpha(w^\nu)}
                                {\partial w^\mu}e_\alpha^a(z^\rho)
\\\nonumber
  &&e^\mu_a(z^\nu) \rightarrow
    \tilde e^\mu_b(w^\nu) = \frac{\partial w^\mu(z^\nu)}
                                 {\partial z^\alpha}e^\alpha_b(z^\rho)
\,.
\end{eqnarray}
The holomorphy of vielbeins,
\begin{eqnarray}\nonumber
    \partial/\partial z^{\bar \mu}e_\nu^a
    =
    1/\sqrt{2}[\partial/\partial{x^{\mu}}+i\partial/\partial y^{\check\mu}]
    [{e_R}_\nu^a + i {e_I}_\nu^a] =0
,
\end{eqnarray}
then implies the Cauchy-Riemann equations,
\begin{equation}
  \frac{\partial{e_R}_\nu^a }{\partial x^{\mu}}
          = \frac{\partial{e_I}_\nu^a }{\partial y^{\check\mu}}
\,,\qquad
  \frac{\partial{e_I}_\nu^a }{\partial x^{\mu}}
         = -\frac{\partial{e_R}_\nu^a }{\partial y^{\check\mu}}
\,.
\label{eq:CauchyRiemann}
\end{equation}
The Cauchy-Riemann equations (\ref{eq:CauchyRiemann}) then imply
that, as a consequence of holomorphy, the tetrads are effectively
functions of {\it four} independent coordinates, even though they
are defined on an {\it eight} dimensional manifold. Thanks to the
holomorphy symmetry, the number of physical degrees of freedom of
our eight dimensional theory is reduced to that of a four
dimensional theory, as required by observations. Conversely, the
knowledge of a complex tetrad (both the real and imaginary parts
of the tetrad must be known) projected onto the $y^{\check\mu}=0$
hypersurface $e_\mu^a(x^\nu,0)$
 plus the holomorphy symmetry allows for the unique reconstruction
of the full eight dimensional dynamics. This feat is achieved by
the simple replacement:
\begin{equation}
e_\mu^a(x^\nu,0)\rightarrow e_\mu^a(\sqrt{2}z^\nu,0)
  \equiv e_\mu^a(z^\nu)
\,. \nonumber
\end{equation}
In this sense our Hermitian gravity is a {\it holographic} theory.
\footnote{
This is of course quite different from 't Hooft's {\it holographic}
principle~\cite{'t Hooft:1993gx} for quantum gravity.
}
Note that this is true only when tetrads are both complex and holomorphic.

Through the inverse relations for the vielbein $e^{\mu}_a
e_{\nu}^a = \delta^{\mu}_{\nu}$ we obtain $
    C^{\mu\bar{\epsilon}}  C_{\bar{\epsilon}\nu}
    =
    \delta^{\mu}_{\nu}
$,
which is in eight dimensional notation equivalent to
\begin{eqnarray}\label{eq:hermitian metric inverse relations}
    \bm C^{me} \bm C_{en}
    =
    \bm \delta^{m}_{n}.
\end{eqnarray}
We can now rotate the line element from
$z^\mu,\bar{z}^{\bar{\mu}}$ space to $x^\mu,y^{\check{\mu}}$ space,
obtaining
\begin{eqnarray}\nonumber
    ds^2
    &=&
    d\boldsymbol{x}^T\cdot \boldsymbol{g} \cdot d\boldsymbol{x}
    =
    d\boldsymbol{x}^m\boldsymbol{g}_{mn}d\boldsymbol{x}^n
   \\
   &=&
        (d x^{\mu},
        d y^{\check{\mu}})
    \left(\begin{array}{cc}
        g_{\mu\nu} & g_{\mu\check{\nu}}\\
        g_{\check{\mu}\nu} & g_{\check{\mu}\check{\nu}}
    \end{array}\right)
    \left(\begin{array}{c}
        d x^{\nu}\\
        d y^{\check{\nu}}
    \end{array}\right),
\label{eq:rotated line element:1}
\end{eqnarray}
where the Latin indices take the values
$0,1,...,d-1,\check{0}$,$\check{1},...,\check{d}-\check{1}$, where
the Greek indices take only the values $0,1,...,d-1$ and where the
number $d$ is again the complex dimension of our manifold. Note
that the equation for the eight dimensional inverse
(\ref{eq:hermitian metric inverse relations}) holds also for the
rotated metric, since its a tensorial equation. We can express the
rotated Hermitian metric components, $\bm g_{mn}$, in terms of
complex metric components, in terms of real and imaginary parts of
the vielbein and in terms of the real symmetric metric
$g_{\mu\nu}$ and real anti-symmetric torsion field $B_{\mu\nu}$,
in the following manner
\begin{eqnarray}\nonumber
    \frac{1}{2}
    \left(\begin{array}{cc}
       (C_{\bar{\mu}\nu}+C_{\mu\bar{\nu}}) & i(C_{\bar{\mu}\nu}-C_{\mu\bar{\nu}})\\[.1cm]
        i(- C_{\bar{\mu}\nu}+C_{\mu\bar{\nu}}) & (C_{\bar{\mu}\nu}+C_{\mu\bar{\nu}})
    \end{array}\right)
     =
     \left(\begin{array}{cc}
        \phantom{-}g_{\mu\nu} & B_{\mu\nu}\\[.1cm]
        -B_{\mu\nu} &  g_{\mu\nu}
    \end{array}\right)
\\\nonumber
=
    \left(\begin{array}{cc}
        ({e_R}^a_{\mu} {e_R}^b_{\nu} + {e_I}^a_{\check{\mu}} {e_I}^b_{\check{\nu}})\eta_{ab} & (-{e_I}^a_{\check{\mu}} {e_R}^b_{\nu} + {e_R}^a_{\mu} {e_I}^b_{\check{\nu}})\eta_{ab}\\[.1cm]
        ({e_I}^a_{\check{\mu}} {e_R}^b_{\nu} - {e_R}^a_{\mu} {e_I}^b_{\check{\nu}})\eta_{ab} & ({e_R}^a_{\mu} {e_R}^b_{\nu} + {e_I}^a_{\check{\mu}}
        {e_I}^b_{\check{\nu}})\eta_{ab}
    \end{array}\right).
\end{eqnarray}
Note that we have defined the eight dimensional rotated Hermitian
metric to be symmetric, $\bm g = \bm g^T$. The rotated Hermitian
line element is then given by
\begin{eqnarray}
    d s^2
    =
    g_{\mu\nu}\left(d x^{\mu}d x^{\nu}
    +
    d y^{\mu}d y^{\nu}\right)
    +
    2B_{\mu\nu}d x^\mu d y^{\nu}.
\label{eq:rotated line element:3}
\end{eqnarray}
Clearly this line element is equal to the Hermitian line element
in its familiar form (\ref{eq:Hermitian familiar form line
element}), when using Einstein's decomposition of the Hermitian
line element,
\begin{eqnarray}\label{eq:Einstein's decomposition of the Hermitian
line element}
    C_{\bar{\mu}\nu}
    =
    g_{\mu\nu} + i B_{\mu\nu},
\end{eqnarray} which basically rotates the
Hermitian line element to $x^\mu,y^{\check{\mu}}$ space
\cite{Einstein:1945eu}. Note that this decomposition exhibits
Hermiticity explicitly, since $g$ is real and symmetric and $B$ is
real and anti-symmetric. The inverse rotated metric can be
expressed in terms of inverse vielbeins in the following manner
\begin{eqnarray}\label{eq:metric inverse g hermitian in vielbeins}
       \boldsymbol{g}^{mn}
        =
    \left(\begin{array}{cc}
        {e_R}^{\mu} {e_R}^{\nu} + {e_I}^{\check{\mu}} {e_I}^{\check{\nu}} & -{e_I}^{\check{\mu}} {e_R}^{\nu} + {e_R}^{\mu} {e_I}^{\check{\nu}}\\[.1cm]
        {e_I}^{\check{\mu}} {e_R}^{\nu} - {e_R}^{\mu} {e_I}^{\check{\nu}} & {e_R}^{\mu} {e_R}^{\nu} + {e_I}^{\check{\mu}} {e_I}^{\check{\nu}}
    \end{array}\right),
\end{eqnarray}
where we have suppressed local indices, $a$ and $b$.

Finally, there is a very handy way of looking at the real and
imaginary parts of the vielbein, which allows us to derive the
just stated objects and their relations in a very trivial manner.
Consider the following holomorphic coordinates $w^\mu(z^{\gamma})
= \frac{1}{\sqrt{2}}(u^{\mu} + i v^{\check{\mu}})$ and
$z^{\nu}(w^{\delta}) = \frac{1}{\sqrt{2}}(x^{\nu} + i
y^{\check{\nu}})$, living in over lapping coordinate patches on a
complex manifold. We can perform a coordinate transformation
\begin{eqnarray}\nonumber
    \left(\begin{array}{c}
       d u^{\mu} \\
       d v^{\check{\mu}}
    \end{array}\right)
    =
    \left(\begin{array}{cc}
        \frac{\partial u^{\mu}}{\partial x^{\nu} } & \frac{\partial u^{\mu}}{\partial y^{\check{\nu}}}\\[.1cm]
        \frac{\partial v^{\check{\mu}}}{\partial x^{\nu} } & \frac{\partial v^{\check{\mu}}}{\partial y^{\check{\nu}}}
    \end{array}\right)
    \left(\begin{array}{c}
        d x^{\nu} \\
        d y^{\check{\nu}}
    \end{array}\right),
\end{eqnarray}
where there are $32$ independent components in the transformation
matrix, because of the Cauchy-Riemann equations
\begin{eqnarray}\nonumber
        \frac{du^{\mu}}{dx^{\nu} }
        =
        \frac{dv^{\check{\mu}}}{dy^{\check{\nu}}}
\phantom{halloda}
        \frac{\partial u^{\mu}}{\partial y^{\check{\nu}}}
        =
        -\frac{\partial v^{\check{\mu}}}{\partial x^{\nu}}.
\end{eqnarray}
We can identify the components of the transformation matrix with
components of vielbeins, when the set $(u,v)$ forms an orthonormal
basis, whenever the $(x,y)$ set forms a coordinate basis. The
Cauchy-Riemann equations in terms of vielbeins then
become\footnote{We can prove this as follows.
$
    \langle\hat{w}^{\mu};\frac{\partial}{\partial\bar{z}^{\bar{\beta}}}\rangle
=
    \frac{1}{2}\langle \hat{u}^{\mu}
    +
    i \hat{v}^{\check{\mu}} ;
    \frac{\partial}{\partial x^{\beta}}
    +
    i\frac{\partial}{\partial y^{\check{\beta}}} \rangle
=
    \frac{1}{2}\langle e^{\mu}_{\alpha}d x^{\alpha}
    +
    e^{\mu}_{\check{\alpha}} d y^{\check{\alpha}}
    +
    i(e^{\check{\mu}}_{\alpha}d x^{\alpha}
    +
    e^{\check{\mu}}_{\check{\alpha}}d y^{\check{\alpha}});
    \frac{\partial}{\partial x^{\beta}}
    +
    i\frac{\partial}{\partial y^{\check{\beta}}}\rangle
=
    \frac{1}{2}[ e^{\mu}_{\beta}
    +
    i( e^{\mu}_{\check{\beta}}
    +
     e^{\check{\mu}}_{\beta})
    -
    e^{\check{\mu}}_{\check{\beta}}].
$ Note that the whole expression is equal to zero by the
definition that vielbeins are holomorphic, because this implies
that $\hat{w}^{\mu} = e^\mu_\alpha (z) dz^{\alpha}$. Now both the
real and imaginary part of the expression vanish, yielding
\begin{eqnarray}\nonumber
        e^{\mu}_{\beta}
        =
        e^{\check{\mu}}_{\check{\beta}}
        \phantom{hallo}
        e^{\mu}_{\check{\beta}}
        = -
        e^{\check{\mu}}_{\beta}.
\end{eqnarray}}
\begin{eqnarray}\label{eq:vielbein Cauchy Riemann relations}
        {e_R}^{\mu}_a
        \equiv
        e^{\mu}_{\alpha}
        =
        e^{\check{\mu}}_{\check{\alpha}}
        \phantom{hallo}
        {e_I}^{\check{\mu}}_a
        \equiv
        e^{\check{\mu}}_{\alpha}
        =
        -e^{\mu}_{\check{\alpha}}.
\end{eqnarray}
These definitions of the real and imaginary parts of the vielbeins
are consistent with the definition of the decomposition of the
complex vielbein in its real and imaginary parts (\ref{eq:viebein
decompositions}). With these definitions the derivation of the
rotated metric tensor is immediate. Since in general relativity
the metric is defined in terms of vielbeins as follows
\begin{eqnarray}\nonumber
        \bm g^{mn} = \bm e^m_k \bm \eta^{kl}\bm e^n_l,
\end{eqnarray}
the $\mu\check{\nu}$ component becomes
\begin{eqnarray}\nonumber
         g^{\check{\mu}\nu}
         =
         e^{\check{\mu}}_k \bm\eta^{kl} e^{\nu}_l
         =
         e^{\check{\mu}}_\kappa \eta^{\kappa\lambda} e^{\nu}_\lambda
         +
         e^{\check{\mu}}_{\check{\kappa}} \eta^{\check{\kappa}\check{\lambda}} e^{\nu}_{\check{\lambda}},
\end{eqnarray}
when defining the rotated flat metric to be $\bm\eta=$
diag$(-1,1,1,1,-1,1,1,1)$. Using the definitions of the real and
imaginary parts of the vielbeins (\ref{eq:vielbein Cauchy Riemann
relations}), the metric component $g^{\check{\mu}\nu}$ is then
given by
\begin{eqnarray}\nonumber
         e^{\check{\mu}}_\kappa \eta^{\kappa\lambda} e^{\nu}_\lambda
         -
         e^{\mu}_{\kappa} \eta^{\kappa\lambda} e^{\check{\nu}}_{\lambda}
         =
          ({e_I}^{\check{\mu}}_a{e_R}^{\nu}_b
          -{e_R}^{\mu}_a{e_I}^{\check{\nu}}_b)\eta^{ab},
\end{eqnarray}
which is consistent with the metric components (\ref{eq:metric
inverse g hermitian in vielbeins}) given earlier. We can now state
the inverse relations for the real and imaginary components of the
vielbeins. In ordinary eight dimensional relativity
\begin{eqnarray}\nonumber
            \bm e^m_k\bm e^k_l = \bm \delta^m_l.
\end{eqnarray}
The $\mu\lambda$ component then becomes
\begin{eqnarray}\label{eq:inverse real and I parts of vielbeins}
            e^\mu_k e^k_\lambda
            =
            \delta^\mu_\lambda
            =
            e^\mu_{\kappa} e^{\kappa}_\lambda
            +
            e^\mu_{\check{\kappa}} e^{\check{\kappa}}_\lambda
            =
            {e_R}^{\mu}_a {e_R}_{\lambda}^a
            +
            {e_I}^{\check{\mu}}_a {e_I}_{\check{\lambda}}^a.
\end{eqnarray}
We can see that $
            C^{\mu\bar{\epsilon}}C_{\bar{\epsilon}\nu}
            =
            g^{\mu\epsilon}g_{\epsilon\nu}+B^{\mu\epsilon}B_{\epsilon\nu},
$ which is equal to the $\mu\nu$ component of $\bm g^{me}\bm
g_{en}$, is indeed equal to $\delta^\mu_\nu$, using the inverse
relations of the real and imaginary parts of the vielbeins
(\ref{eq:inverse real and I parts of vielbeins}).

\section{Flat Space}
\noindent The Hermitian line element in flat space becomes
\begin{eqnarray}
    ds^2
    =
    -(c d t)^2 + (d\vec{x})^2
    -(dy^0)^2 + (d\vec{y})^2 .
\label{eq:line element:dHermitian}
\end{eqnarray}
From now on we declare the $y$ coordinate to be the
energy-momentum coordinate by defining\footnote{In principle one
can pick any constant such that the units come out right. If one
decides to construct the constant without introducing new
constants, Newton's constant over the speed of light cubed is a
unique choice up to factors of order unity. Since we are, up to
this point, constructing a classical theory, there is no room for
Planck's constant.} $y^\mu \equiv p^\mu\frac{G_N}{c^3}$. The
space-time-momentum-energy interval squared from the origin to a
space-time-momentum-energy point $\bm x =
(ct,\vec{x},\frac{G_N}{c^4}E,\frac{G_N}{c^3}\vec{p})$ is given by
\begin{align}\label{eq:linelement spacetimemomentum energy flatspace her on-shell }
    d^2\left(\bm 0;\bm x\right)
    =
    -(ct)^2 + (\vec{x})^2 +
    \frac{G_N^2}{c^6}
    \left[ (\vec{p})^2-\left(\frac{E}{c}\right)^2 \right],
\end{align}
where ${G_N^2}/{c^6} $ suppresses the momentum-energy part by
a factor on the order of $10^{-72} {s^2}/{kg^2}$, as it should
do, since we do not observe any momentum-energy contributions
at low energies.

\begin{figure}
  \centering
  \includegraphics[width=\columnwidth]{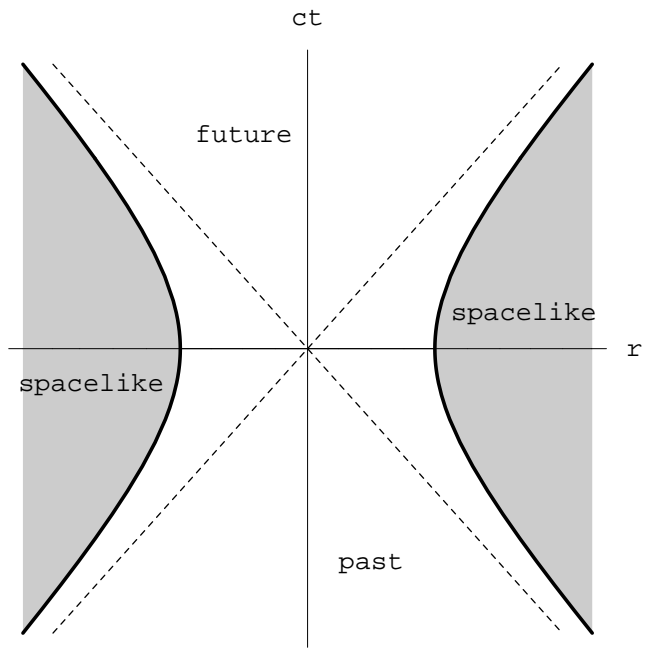}
  \caption{A light cone, modified by non inertial coordinate transformations,
  is being portrayed on a space-time-momentum-energy diagram, separating the
  regions that are in causal contact with each other,
  from the regions that are not. There is a nonlocal causally related volume element at the origin.}
  \label{fig:LightConeher}
\end{figure}
The group of transformations, which leaves the Hermitian metric
(\ref{eq:Hermitian familiar form line element}) invariant, is the
$U(1,3)$ group; the elements $U$ of the $U(1,3)$ group satisfy by
definition the relation $U^\dagger \eta U = \eta $. When
considering only one space and one momentum dimension, instead of
three of each, we can represent the elements of the $SU(1,1)$
group~\cite{Low:2005tb},
operating on the space-time-momentum-energy vector $(t,x,p,E)$, by
\begin{eqnarray}\nonumber
   \Gamma(v,f,f_0)
   =
   \gamma(v,f,f_0)
   \left(
\begin{array}{cccc}
   1        & \frac{v}{c^2}         & \frac{G_N^2 f}{c^8}       & -\frac{G_N^2 f_0}{c^{9}} \\[0.1cm]
   v        &    1                  & \frac{G_N^2 f_0}{c^7}     & -\frac{G_N^2 f}{c^8}      \\
   f        & -\frac{f_0}{c}      & 1                         & \frac{v}{c^2}             \\
   c f_0    & -f                    & v                         & 1                         \\
   \end{array}
   \right),
\end{eqnarray}
where $ \gamma(v,f,f_0) = \left(1 - \frac{v^2}{c^2} - \frac{G_N^2
f^2}{c^8} + \frac{G_N^2 f_0^2}{c^{8}}\right)^{-\frac{1}{2}}$ and
where $f = \dot{p}$ and $f_0 = \frac{\dot{E}}{c}$. We have
neglected two space and two momentum components for convenience,
since the transformations between the spacial components are
simply rotations in space. An overall phase factor can be added
later if one wants to consider the $U(1,1)$ group instead of the
$SU(1,1)$ group. Note that the $SU(1,3)$ group reduces to its
subgroup the Lorentz group, the $SO(1,1)$ group, for $f_i$ and
$f_0$ being zero. Thus, for inertial frames the theory clearly
reduces to the laws of special relativity.

If we set $ds^2 = 0$ and there is no momentum-energy contribution,
we know that $\dot{\vec{x}}= c$. Demanding again that $ds^2 = 0$,
but now for non vanishing momentum energy contributions however,
the space-time-momentum-energy interval (\ref{eq:linelement
spacetimemomentum energy flatspace her on-shell }) becomes
\begin{eqnarray}\label{eq:distance flatspace her on-shell}
-(ct)^2 + (\vec{x})^2 + \frac{G_N^2 p^2}{c^6} = 0,
\end{eqnarray}
where $ p^2 = p^\mu p_\mu $ measures the energy-momentum in a
space-time-momentum-energy hyper surface. Setting the
space-time-momentum-energy interval to zero determines the
causality boundary: the hypersurface specified by this condition
determines the boundary of the causally related regions
(for causally related events the `$=$' sign
in~(\ref{eq:distance flatspace her on-shell})
should be replaced by a `$\leq$').

 Consider now the case when the momentum-energy contribution $p^2$ is
negative. We can furthermore restrict the  hypersurface
by setting $t$ to zero. Solving for $\vec{x}$ we obtain
\begin{eqnarray}\nonumber
    r_{\rm max}
    =
    \frac{G_N\sqrt{- p^2}}{c^3} ,
\end{eqnarray}
where $r_{\rm max} = \sqrt{\vec x^2(t=0)}$ is the maximal radius
for a spatial volume element which is simultaneously causally
related in a nonlocal manner. This can be seen from figure
(\ref{fig:LightConeher}), which depicts this causality boundary.
The existence of a momentum dependent maximal radius (often
referred to as a minimal length scale) suggests that nature is
inherently nonlocal for non-inertial observers. When $p^2$ is zero
the standard light-cone is recovered. To get a feeling on how big
this `violation of causality' can be, let us consider a particle
on the momentum-energy shell (corresponding to the traditional
{\it on-shell} notion), in which case $p^2 = -m^2c^2$, such that
$r_{\rm max} = G_Nm/c^2 = r_{\rm Sch}/2$, where $r_{\rm Sch}$
denotes the conventional Schwarzschild radius. Recall that our
flat-space analysis is based on the geodesic equation and its
integral the line element, and hence completely neglects the
self-gravity of (elementary) particles. When the self-gravity
effects are included however, we expect that the above-discussed
violation of causality gets hidden by the Schwarzschild radius
created by the particles in consideration. It would be of interest
to consider in detail how the causality analysis gets modified
when the self gravity of particles is taken into account.

Alternatively, we can write $r_{\rm max} = G_Nm/c^2 = (m/m_{\rm
Pl})\ell_{\rm Pl}$, where we introduced the conventional Planck
mass $m_{\rm Pl} = \sqrt{c\hbar/G_N}\simeq 2.18\times 10^{-8}{\rm
kg}$ and the Planck length $\ell_{\rm Pl}=\sqrt{\hbar
G_N/c^3}\simeq 1.616\times 10^{-35}{\rm m}$, and where $\hbar$
denotes the Planck constant. Even though $r_{\rm max}$ is a purely
classical quantity, when represented in terms of the Planck units,
the Planck constant appears (which gets, of course, cancelled in
the ratio $\ell_{\rm Pl}/m_{\rm Pl}$).

\begin{figure}
  \centering
  \includegraphics[width=\columnwidth]{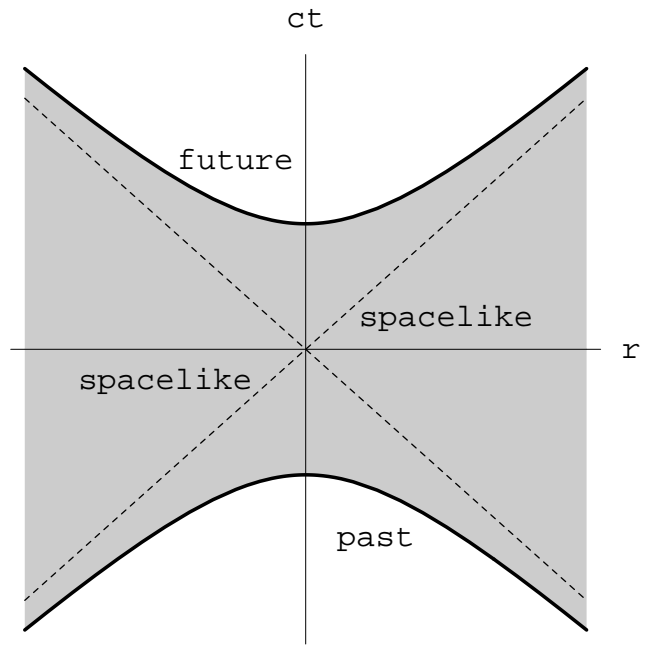}
  \caption{A light cone, modified by non inertial coordinate transformations,
  is being portrayed on a space-time-momentum-energy diagram separating the regions that are in causal contact each other,
  from the regions that are not. There is a minimal time interval
  for events to be in causal contact.}
  \label{fig:LightConeHer1}
\end{figure}

We can also specify our hypersurface differently by setting
$\vec{x}$ to zero instead of $t$ (this is in the case of positive
momentum-energy, $p^2$). Solving for $t$ we obtain
\begin{eqnarray}\nonumber
t_{\rm min} =  \frac{G_N\sqrt{p^2}}{c^4} ,
\end{eqnarray}
where $t_{\rm min}$ is the minimal time for which an event is
causally related to past space-time-momentum-energy events, as can
be seen from the space-time-momentum-energy diagram,
figure~\ref{fig:LightConeHer1}. This means that $t_{\rm min}$ is
the minimal time for which an event can influence future events.
Equivalently, $t_{\rm min}$ is the minimal time for which an event
can be influenced by past events. In the light of the above
discussion of $r_{\rm max}$, we see that the minimal time $t_{\rm
min}$ shown in figure~\ref{fig:LightConeHer1} exists only for
particles moving off-shell, and hence these particles are strictly
speaking not classical.

The definitions of space-time-momentum-energy past and future are
natural generalizations of space-time past and future; the regions
that are in causal contact are given again by events removed from
the origin by space-time-momentum-energy intervals smaller or
equal then zero, $d s^2 \leq 0$. We have thus shown that in the
limit of weak fields, Hermitian gravity exhibits simultaneously
connected regions (maximal distance) for time like energy-momentum
intervals and a breach of causality (minimal time) for space-like
frame energy-momenta. For light-like frame energy-momenta,
Hermitian gravity reproduces the standard light causality
structure characterizing the weak field limit of general
relativity.

We now calculate the phase velocity, using the
space-time-momentum-energy line element (\ref{eq:distance
flatspace her on-shell})
\begin{eqnarray}\label{eq:phase velocity hermitian case}
    v_{\rm phase}
    =
    \frac{\|\vec r\,\|}{t}
    = \sqrt{c^2  - \frac{G_N^2}{c^6}\frac{p^2}{t^2}}
\,.
\end{eqnarray}
One can see that 
the phase velocity approaches the conventional speed of light $c$
for large $t$. For times smaller than the minimal time the phase
velocity becomes imaginary and damping will occur. The group
velocity for massless particles (${d s}/{dt} = 0$) becomes
\begin{eqnarray}\label{eq:groupvelocity hermtian case}
    v_{g}
    =
    \sqrt{c^2 - \frac{G_N^2}{c^6}f^2},
\end{eqnarray}
where the four force squared is given by
\begin{eqnarray}
    f^2
    =
    -\frac{1}{c^2}\left(\frac{d E}{d t}\right)^2
    +
    \left(\frac{d\vec{p}}{d t}\right)^2.
\label{eq:four force2}
\end{eqnarray}
The maximal group velocity approaches the speed of light for small
four forces squared $f^2$. The maximal group velocity must be real
in order to facilitate propagation. The reality requirement
implies
 $f^2\leq {c^4}/{G_N}$. This can be seen from
the space-time-momentum-energy line element~(\ref{eq:line element:dHermitian})
divided by $(dt)^2$.
Upon solving~(\ref{eq:line element:dHermitian}) for the four force
squared~(\ref{eq:four force2}),
we obtain $({G_N^2}/{c^6}) f^2
\equiv ({G_N^2}/{c^6})\left({d p}/{d t}\right)^2 =
c^2 - v_g^2 + \left({d s}/{d t}\right)^2 \leq c^2 - v_g^2\leq c^2$,
where the inequalities follow from the observations that
$(ds/dt)^2\leq 0$ and $v_g^2\geq 0$
(the reality condition on $v_g$ in Eq.~(\ref{eq:groupvelocity hermtian case})).
This then implies
\begin{eqnarray}\nonumber
    f^2
    \leq f_{\rm max}^2
     = \frac{c^8}{G_N^2}.
\end{eqnarray}
There is no lower bound on $f^2$, since $\left({d s}/{d
t}\right)^2\leq 0$ can at least in principle be arbitrarily large
and negative. This means that there is also no upper bound on the
maximal group velocity~(\ref{eq:groupvelocity hermtian case}), as
$f^2$ can in principle be very large and negative. While this may
be true in principle, more realistically -- in the flat space
limit and in the absence of external forces -- the flat space
geodesic equation implies that $dz^\mu/d\tau = U^\mu = ({\rm
constant})^\mu$, from which we conclude that $f^2$ must be {\rm
constant}, such that the (maximal allowed) group
velocity~(\ref{eq:groupvelocity hermtian case}) acquires a {\it
constant} correction in flat spaces
 and in the absence of external forces.
This type of corrections can play an important role in strongly curved
space-times however, where strong gravitational forces
exist, which  are expected to induce large changes in $f^2$,
and thus possibly {\it superluminal} propagation.
The change of causality structure discussed in this section
deserves a deeper analysis, since it
might have important consequences for the physics of structure
formation in the early universe~\cite{Moffat:1992ud,Barrow:1999jq}.


\section{The Equations of Motion}

\noindent In order to write down the equations of motion for this
Hermitian theory of gravity, one needs to know what the connection
coefficients are. There are already connection coefficients which
are called Hermitian connection coefficients~\cite{Nakahara:2003nw}.
We will derive different connection coefficients later, but in order
to appreciate these newly obtained connection coefficients, we
will consider the old ones first.

\subsection{The Known Connection Coefficients}

\noindent One can derive the known Hermitian connection
coefficients easily if one requires metric compatibility of the
Hermitian metric and the fact that the holomorphic covariant
derivative of an anti-holomorphic basis vector vanishes. The
vanishing of he holomorphic covariant derivative of an
anti-holomorphic basis vector is given by
\begin{eqnarray}\label{eq:covariant derivative known}
    \nabla_\mu \frac{\partial}{\partial z^{\bar{\nu}}}
    =
    0
    \phantom{halloda}
    \nabla_{\bar{\mu}} \frac{\partial}{\partial z^{\nu}}
    =
    0.
\end{eqnarray}
This implies that $\Gamma(\text{mixed indices}) = 0$, since the
complex connection coefficients are usually defined as
\begin{eqnarray}\nonumber
    \nabla_\mu \frac{\partial}{\partial z^{\bar{\nu}}}
    =
    \Gamma^{\epsilon}_{\mu\bar{\nu}} \frac{\partial}{\partial z^\epsilon}
    =
    0.
\end{eqnarray}
If one then imposes metric compatibility on the Hermitian metric
\begin{subequations}\label{eq:vielbein compatibilty}
\begin{eqnarray}\label{eq:vielbein compatibilty a}
    \nabla_\rho C_{\bar{\mu}\nu}
    =
    \partial_\rho C_{\bar{\mu}\nu}
    -
    C_{\bar{\mu}\lambda}\Gamma^\lambda_{\rho\nu}
    =
    0
\end{eqnarray}
\begin{eqnarray}\label{eq:vielbein compatibilty b}
    \nabla_{\bar{\rho}} C_{\bar{\mu}\nu}
    =
    \partial_{\bar{\rho}} C_{\bar{\mu}\nu}
    -
    C_{\bar{\lambda}\nu}\Gamma^{\bar{\lambda}}_{\bar{\rho}\bar{\mu}}
    =
    0
\end{eqnarray}
\end{subequations}
one can easily read off the Hermitian connection coefficients
\begin{eqnarray}\label{eq:connection known}
    \Gamma^\lambda_{\rho\nu}
    =
    C^{\bar{\epsilon}\lambda}\partial_\rho C_{\nu\bar{\epsilon}}
    \phantom{halloda}
    \Gamma^{\bar{\lambda}}_{\bar{\rho}\bar{\nu}}
    =
    C^{\bar{\lambda}\epsilon}\partial_{\bar{\rho}}
    C_{\epsilon\bar{\nu}}.
\end{eqnarray}
If one looks carefully at the Hermitian metric compatibility
equations (\ref{eq:vielbein compatibilty}) one can easily see that
these equations imply vielbein compatibility; one can obtain the
Hermitian connection coefficients by imposing vielbein
compatibility in the following manner
\begin{eqnarray}\label{eq:vielbein compat}
    \nabla_\mu e_\nu
    =
    0
    \phantom{ha}
    \nabla_{\bar{\mu}} e_\nu
    =
    0
    \phantom{ha}
    \nabla_\mu e_{\bar{\nu}}
    =
    0
    \phantom{ha}
    \nabla_{\bar{\mu}} e_{\bar{\nu}}
    =
    0.
\end{eqnarray}
Theories of Hermitian gravity, satisfying vielbein compatibility,
have been proposed \cite{Chamseddine:2005at}. We shall see below
that in such theories the geodesic equation is not obtained via an
action principle. In an attempt to fix this problem we will weaken
this vielbein compatibility condition and obtain different
connection coefficients.

Note that the two independent components of the Riemann tensor,
\begin{eqnarray}\label{eq:Riemann known}
    &&R^{\kappa}_{\lambda\bar{\mu}\nu}
    =
    \partial_{\bar{\mu}}\Gamma^{\kappa}_{\nu\lambda}
    =
    \partial_{\bar{\mu}}\left(C^{\bar{\epsilon}\kappa}\partial_{\nu}
         C_{\lambda\bar{\epsilon}}\right)
    \\\nonumber
    &&R^{\bar{\kappa}}_{\bar{\lambda}\mu\bar{\nu}}
    =
    \partial_{\mu}\Gamma^{\bar{\kappa}}_{\bar{\nu}\bar{\lambda}}
    =
    \partial_{\mu}
     \left(C^{\bar{\kappa}\epsilon}\partial_{\bar{\nu}}
             C_{\epsilon\bar{\lambda}}\right)
\,,
\end{eqnarray}
contain only first order derivatives, when assuming the Hermitian
metric to be a product of a holomorphic and anti-holomorphic
vielbein $C_{\bar{\mu}\nu} = e_{\bar{\mu}}^a (\bar{z}^\gamma)
\eta_{ab}e_{\nu}^b (z^\gamma)$. This implies that if one would
attempt to write a complex equation of motion, analogues to
Einstein's equations, one would obtain a first order differential
equation. This means that the space-time-momentum-energy curvature
for this theory is non-dynamical.

\subsection{The Hermitian Geodesic Equations}

\noindent The easiest way to derive the connection coefficients
for Hermitian gravity is through varying the Hermitian line
element. One then obtains the Hermitian geodesic equations from
which one can read off the connection coefficients. Varying the
Hermitian line element is a very easy exercise, when considering
its eight dimensional form (\ref{eq:line element hermitian metric
subsection}). One then obtains the eight dimensional complex
geodesic equations
\begin{eqnarray}\label{eq:geodesic equation arbitrary complex}
    \boldsymbol{\ddot{z}}^r
  + \boldsymbol{\Gamma}^r_{mn} \boldsymbol{\dot{z}}^m \boldsymbol{\dot{z}}^n
    = \boldsymbol{0},
\end{eqnarray}
where the complex connection coefficients are given by
\begin{eqnarray}\label{eq:connection coefficients arbitrary complex}
   \boldsymbol{ \Gamma}^r_{mn}
   =
   \frac{1}{2}\boldsymbol{C}^{re}(\bm\partial_m \boldsymbol{C}_{en}
      + \bm\partial_n \boldsymbol{C}_{me}-\bm\partial_e
   \boldsymbol{C}_{mn}).
\end{eqnarray}
The Hermitian metric is defined such that the $\mu\bar{\nu}$
component of ${\bm C}_{mn}$ is $C_{\mu\bar{\nu}}$, with vanishing
unmixed components. The Hermitian geodesic equations are then
given by
\begin{eqnarray}\nonumber
    \ddot{z}^\rho
    +
    \Gamma^\rho_{\mu\nu}\dot{z}^{\mu}\dot{z}^{\nu}
    +
    \Gamma^\rho_{\bar{\mu}\nu}\dot{z}^{\bar{\mu}}\dot{z}^{\nu}
    +
    \Gamma^\rho_{\mu\bar{\nu}}\dot{z}^{\mu}\dot{z}^{\bar{\nu}}
    +
    \Gamma^\rho_{\bar{\mu}\bar{\nu}}\dot{z}^{\bar{\mu}}\dot{z}^{\bar{\nu}}
    &=&
    0
    \\\nonumber
    \ddot{z}^{\bar{\rho}}
    +
    \Gamma^{\bar{\rho}}_{\mu\nu}\dot{z}^{\mu}\dot{z}^{\nu}
    +
    \Gamma^{\bar{\rho}}_{\bar{\mu}\nu}\dot{z}^{\bar{\mu}}\dot{z}^{\nu}
    +
    \Gamma^{\bar{\rho}}_{\mu\bar{\nu}}\dot{z}^{\mu}\dot{z}^{\bar{\nu}}
    +
    \Gamma^{\bar{\rho}}_{\bar{\mu}\bar{\nu}}\dot{z}^{\bar{\mu}}\dot{z}^{\bar{\nu}}
    &=&
    0,
\end{eqnarray}
where the connection coefficients are
\begin{align}\label{eq:Hermitian connection coefficients z basis 4 d}
    \Gamma^\rho_{\mu\nu}
    &=
    \frac{1}{2}C^{\bar{\lambda}\rho}(\partial_{\mu} C_{\nu\bar{\lambda}} + \partial_{\nu} C_{\mu\bar{\lambda}})
    \\\nonumber
    \Gamma^\rho_{\bar{\mu}\nu}
    &=
    \frac{1}{2}C^{\bar{\lambda}\rho}(\partial_{\bar{\mu}} C_{\nu\bar{\lambda}} - \partial_{\bar{\lambda}} C_{\nu\bar{\mu}})
    \\\nonumber
    \Gamma^\rho_{\mu\bar{\nu}}
    &=
    \frac{1}{2}C^{\bar{\lambda}\rho}(\partial_{\bar{\nu}} C_{\mu\bar{\lambda}} - \partial_{\bar{\lambda}} C_{\mu\bar{\nu}})
    \\\nonumber
    \Gamma^{\bar{\rho}}_{\bar{\mu}\bar{\nu}}
    &=
    \frac{1}{2}C^{\bar{\rho}\lambda}(\partial_{\bar{\mu}} C_{\lambda\bar{\nu}} + \partial_{\bar{\nu}} C_{\lambda\bar{\mu}})
    \\\nonumber
    \Gamma^{\bar{\rho}}_{\mu\bar{\nu}}
    &=
    \frac{1}{2}C^{\bar{\rho}\lambda}(\partial_{\mu} C_{\lambda\bar{\nu}} - \partial_{\lambda} C_{\mu\bar{\nu}})
    \\\nonumber
    \Gamma^{\bar{\rho}}_{\bar{\mu}\nu}
    &=
    \frac{1}{2}C^{\bar{\rho}\lambda}(\partial_{\nu} C_{\lambda\bar{\mu}} - \partial_{\lambda} C_{\nu\bar{\mu}})
    \\\nonumber
     \Gamma^\rho_{\bar{\mu}\bar{\nu}}
    &=
    0\,,
    \phantom{halloda}
    \Gamma^{\bar{\rho}}_{\mu\nu}
    =
    0.
\end{align}
These connection coefficients are Hermitian in the following
sense, $\overline{\Gamma^{\rho}_{\mu\bar{\nu}}} =
{\Gamma^{\bar{\rho}}_{\bar{\mu}\nu}}^T$ and
$\overline{\Gamma^{\rho}_{\mu\nu}} =
{\Gamma^{\bar{\rho}}_{\bar{\mu}\bar{\nu}}}^T$. In eight
dimensional form the connection coefficients are symmetric, $
({\bm \Gamma^{r}_{mn}})= ({\bm\Gamma^{r}_{nm}})$, just as the
Levi-Civita symbols in general relativity. The wisdom of the eight
dimensional notation becomes apparent now; for any known equation
of general relativity one can sum over the barred and unbarred
indices and plug in the just derived connection coefficients. If
one plugs in the connection coefficients (\ref{eq:Hermitian
connection coefficients z basis 4 d}) into the Hermitian geodesic
equations one obtains
\begin{align}\nonumber
    \ddot{z}^\rho
    +
    \frac{1}{2}&C^{\rho\bar{\lambda}}
    (\partial_{\mu} C_{\nu\bar{\lambda}} + \partial_{\nu} C_{\bar{\lambda}\mu})
    \dot{z}^{\mu}\dot{z}^{\nu}
    +
    \\\nonumber
    &C^{\rho\bar{\lambda}}
    (\partial_{\bar{\nu}} C_{\mu\bar{\lambda}} - \partial_{\bar{\lambda}} C_{\bar{\nu}\mu})
    \dot{z}^{\mu}\dot{z}^{\bar{\nu}}
    =
    0
\end{align}
and its Hermitian conjugate. One obtains precisely this
result, when varying the particle action (with a mass $m$),
\begin{equation}
S=-m\int ds \nonumber
\end{equation}
with respect to the real one dimensional parameter proper time
$\tau$, where $ds$ represents the Hermitian line
element~(\ref{eq:Hermitian familiar form line element}). Note that
the known Hermitian connection coefficients (\ref{eq:connection
known}) cannot be derived from any variation principle and
therefore, in that sense, cannot have any physical meaning.

From general relativity we know that the metric transforms as
\begin{align}\nonumber
    \bm {C}_{mn}
    \rightarrow
    \bm C_{mn}
    +
    \bm \nabla_{m}\bm\Lambda_{n}
    +
    \bm \nabla_{n}\bm\Lambda_{m}.
\end{align}
One can check that for example the $\mu\bar{\nu}$ component of
this equation with the connection coefficients (\ref{eq:Hermitian
connection coefficients z basis 4 d}) plugged in corresponds
indeed to the transformation of the Hermitian metric
$C_{\mu\bar{\nu}}$, where $z^{\mu}\rightarrow z^{\mu} +
\Lambda^{\mu}$ and $C_{\mu\bar{\nu}} \rightarrow C_{\mu\bar{\nu}}
+ \Lambda^{\epsilon}\partial_{\epsilon}C_{\mu\bar{\nu}}+
\Lambda^{\bar{\epsilon}}\partial_{\bar{\epsilon}}C_{\mu\bar{\nu}}$.
This should strengthen our belief in the connection coefficients
(\ref{eq:Hermitian connection coefficients z basis 4 d}). Note
that via this procedure we can only obtain the coefficients with
mixed indices. The unmixed coefficients can be obtained from the
variation of the particle action with respect to the real variable
proper time, because this variable breaks homomorphy in the sense
that it depends on both $z^{\mu}$ and $z^{\bar{\mu}}$. 

Finally, since we can simply rotate any tensorial equation from
$z^{\mu},z^{\bar{\mu}}$ space to $x^{\mu},y^{\check{\mu}} $ space,
it is useful to have expressions for the rotated connection
coefficients, such that we can just plug these rotated
coefficients into the rotated equations. Although the connection
transforms by definition as a connection, it clearly transforms as
a $(1,2)$ tensor under these rotations. This is the case, because
these rotations are just constant transformations. Hence we can
just replace the complex metric, $\bm C_{mn}$, by the rotated
metric, $\bm g_{mn}$, in the expression of the eight dimensional
connection (\ref{eq:connection coefficients arbitrary complex})
\cite{MantzThesis:2007}. We state two components of the rotated
connection coefficients
\begin{align}\nonumber
    \Gamma^{\rho}_{\mu\nu}
    =
    \frac{1}{2}g^{\rho\epsilon}
    (\partial_{\mu} g_{\epsilon\nu} &+ \partial_{\nu} g_{\mu\epsilon} -\partial_{\epsilon} g_{\mu\nu})
    \\\nonumber
    +
     \frac{1}{2}B^{\rho\check{\epsilon}}
    (\partial_{\mu} B_{\check{\epsilon}\nu} &+ \partial_{\nu} B_{\mu\check{\epsilon}} -\partial_{\check{\epsilon}} g_{\mu\nu})
\\\nonumber
    \Gamma^{\rho}_{\check{\mu}\nu}
    =
    \frac{1}{2}g^{\rho\epsilon}
    (\partial_{\check{\mu}} g_{\epsilon\nu} &+ \partial_{\nu} B_{\check{\mu}\epsilon} -\partial_{\epsilon} B_{\check{\mu}\nu})
    \\\nonumber
    +
     \frac{1}{2}B^{\rho\check{\epsilon}}
    (\partial_{\check{\mu}} B_{\check{\epsilon}\nu} &+ \partial_{\nu} g_{\check{\mu}\check{\epsilon}} -\partial_{\check{\epsilon}}
    B_{\check{\mu}\nu}).
\end{align}

\subsection{Torsion and Curvature}

The Hermitian torsion\footnote{In this article we define dynamical
torsion to be a second order differential equations constraining
the anti-symmetric part of the metric (\ref{eq:Einstein's
decomposition of the Hermitian line element}). This has in
principle nothing to do with the vanishing or nonvanishing of the
torsion tensor.} tensor $T$ and the Riemann tensor $R$ are defined
as
\begin{eqnarray}\nonumber
    T(Z,W)
    =
    \nabla_{Z}W - \nabla_{W}Z - [Z,W]
    \\\nonumber
    R(Z,W)V
    =
    \nabla_{Z} \nabla_{W}V -  \nabla_{W} \nabla_{Z}V -
    \nabla_{[Z,W]}.
\end{eqnarray}
The covariant derivative acting on a basis vector yields
\begin{eqnarray}\nonumber
    \bm \nabla_m \frac{\bm\partial}{\bm\partial \bm z^n}
    =
    \bm\Gamma^{e}_{m n}\frac{\bm\partial}{\bm\partial \bm z^e}.
\end{eqnarray}
Note that the $\mu\bar{\nu}$ component of the covariant derivative
acting on the basis vector is this time non-vanishing, unlike the
$\mu\bar{\nu}$ component of the covariant derivative acting on the
basis vector (\ref{eq:covariant derivative known}), using the
known coefficients (\ref{eq:connection known}). We can now write
the expressions for the components of the Hermitian torsion tensor
\begin{eqnarray}\nonumber
    \bm T^{l}_{\phantom{l}mn}
    &=&
     \langle  {\hat{\bm e}}^l,
          {\bm T}({\hat{\bm e}}_{m},{\hat{\bm e}}_{n})\rangle
    =
    \langle{\hat{\bm e}}^l, {\bm\nabla_m} {\hat{\bm e}}_{n}
       - {\bm\nabla_n} {\hat{\bm e}}_m
\rangle
    \\\nonumber
    &=&
\langle
 {\hat{\bm e}}^l,
{\bm\Gamma^{b}_{mn}}{\hat{\bm e}}_b
      - {\bm\Gamma^{b}_{nm}}{\hat{\bm e}}_b
\rangle
    =
    {\bm \Gamma^{l}_{mn}} - {\bm\Gamma^{l}_{nm}}
    =
    0
\end{eqnarray}
and the Hermitian Riemann tensor
\begin{eqnarray}\nonumber
    {\bm R^{s}_{mln}} =
    {\bm\partial_{l}}{\bm\Gamma^{s}_{nm}} -
    {\bm\partial_{n}}{\bm\Gamma^{s}_{lm}}
    + {\bm\Gamma^{s}_{la}}{\bm\Gamma^{a}_{nm}} -
    {\bm\Gamma^{s}_{na}}{\bm\Gamma^{a}_{lm}},
\end{eqnarray}
with the connection coefficients (\ref{eq:Hermitian connection
coefficients z basis 4 d}) plugged in. Hence the Hermitian Riemann
tensor is Hermitian in the following sense
$\overline{R_{\bar{\mu}\nu}} = {R_{\mu\bar{\nu}}}^T $. Therefore
the Hermitian Ricci scalar is real, $ \overline{R} = R$.

\subsection{Action principle for Hermitian gravity}

\noindent The action for Hermitian gravity can be formulated as,
\begin{eqnarray}
   S[{\bm C},\psi_i] = S_{hg}[{\bm C}]+S_{c}[{\bm C}]
                       + S_{M}[{\bm C},\psi_i]
\label{eq:complex action:full}
\end{eqnarray}
where the pure gravity action is the following generalization of
the Hilbert-Einstein action,
\begin{eqnarray}\nonumber
   S_{hg}[{\bm C}]
   = \frac{1}{16\pi G_N}\int d {\bm z}^8 \sqrt{\bm {\bm C}} ({\bm R}-2\Lambda)
\, ,
\end{eqnarray}
where $
    {\bm R}
    =
    {\bm C^{mn}}{\bm R_{mn}}
$ denotes the Ricci scalar, $\Lambda$ cosmological constant and
${\bm C^{mn}}$ denotes the full complex metric
tensor~(\ref{complex metric}). For the reasons explained below, we
impose the reciprocity symmetry only at the level of equations of
motion (on-shell), which at the level of the action can be
realized by a constraint. This of course means that physical
quantities still respect the reciprocity symmetry. There is no
unique way of imposing the reciprocity symmetry on the metric
tensor. One reasonable choice is the following `particle' action,
\begin{eqnarray}\label{eq:constraint action}
    S_{c}[{\bm C}]
=
    - M \int \Big[\lambda_1(C_{\mu\nu}dz^\mu dz^\nu
    &+&
    C_{\bar\mu\bar\nu}dz^{\bar\mu} dz^{\bar\nu})
\\\nonumber
    + \,\lambda_2(C_{\mu\bar\nu}dz^\mu dz^{\bar\nu}
    &+&
    C_{\bar\mu\nu}dz^{\bar\mu} dz^\nu)
              \Big]^{1/2},
\end{eqnarray}
where $M$ is a (mass) parameter and $\lambda_1$, $\lambda_2$ are
Lagrange multipliers which break the symmetry between the
holomorphic and Hermitian components of the complex
metric~(\ref{complex metric}). Taking, for example,
$\lambda_1=\lambda$ and $\lambda_2=1$, imposes
$C_{\mu\nu}=0=C_{\bar\mu\bar\nu}$ and thus on-shell Hermiticity of
the metric. Conversely, when $\lambda_1= 1$ and $\lambda_2=
\lambda$ imposes $C_{\mu\bar\nu}=0=C_{\bar\mu\nu}$, implying
on-shell holomorphy of the metric tensor. The constraint
action~(\ref{eq:constraint action}) does not break holomorphy of
the full theory~(\ref{eq:complex action:full}) realized at the
level of tetrads.

 Just like the gravitational action, which obeys holomorphy at the level
of vielbeins, we shall require that the matter action
in~(\ref{eq:complex action:full}) consists of holomorphic matter fields.
Namely, holomorphy reduces the large number of degrees of freedom of the full
{\it eight} dimensional theory to an acceptable number of degrees of freedom
of an effectively {\it four} dimensional theory, as observationally required.
For simplicity here we consider a matter action for scalar fields, which
we use extensively below when we study cosmology.
We consider two holomorphic scalar fields $\phi$ and $\psi$,
one with Hermitian and one with holomorphic kinetic term, with the action:
\begin{eqnarray}\label{eq:scalar action}
 S_M[\phi,\psi] = \int d^8{\bm z}\sqrt{{\bm C}}\, {\cal L},
\end{eqnarray}
where the lagrangian density is given by
\begin{eqnarray}\nonumber
 {\cal L} =  - \frac{\alpha}{2}{\bm C}^{mn}(\partial_m\Phi)^\dagger
                    \cdot\partial_n\Phi
          - \frac{\beta}{2}{\bm C}^{mn}(\partial_m\Psi)^T\cdot\partial_n\Psi
          - V
\end{eqnarray}
where
\begin{equation}
\Phi  = \left({\phi\atop{\bar\phi}}\right)
\,,\qquad
\Psi  = \left({\psi\atop{\bar\psi}}\right)
\,,
\label{eq:scalar action:fields}
\end{equation}
where $\alpha$ and $\beta$ are constants and where
$V=V(\Phi,\Psi)$ is a potential. Note that $\Phi^\dagger\cdot\Phi
=  2\phi\bar\phi$ and
 $\Psi^T\cdot\Psi =  \psi^2+\bar\psi^2$.
The constants $\alpha$ and $\beta$ can be absorbed in the fields
$\phi$ and $\psi$ by the appropriate field redefinitions, except for
the sign of $\alpha$, which is an invariant and thus
can have physical relevance.
 For simplicity, we have assumed
in Eq.~(\ref{eq:scalar action}) that the scalar fields
do not couple to the Ricci scalar.

Varying the action~(\ref{eq:complex action:full}) results in
the Hermitian Einstein-Hilbert equations of motion
\begin{eqnarray}
  {\bm G}_{mn} + \Lambda {\bm C}_{mn} &=& 8\pi G_N {\bm T}_{mn}
\nonumber\\
     C_{\mu\nu} &=& 0 = C_{\bar\mu\bar\nu}
\,,
\label{eq:hermitian eom}
\end{eqnarray}
where the second line equation is obtained by choosing $\lambda_1=\lambda$,
$\lambda_2=1$ and varying the action~(\ref{eq:constraint action})
 with respect to $\lambda$.
As usual the following definitions hold for the Einstein tensor
${\bm G}_{mn}$ and the stress energy tensor ${\bm T}_{mn}$:
\begin{eqnarray}
    {\bm G}_{mn} &=& {\bm R}_{mn} - \frac 12 {\bm C}_{mn}{\bm R}
\,,\qquad    {\bm R} = {\bm C}^{mn}{\bm R}_{mn}
\nonumber\\
    {\bm T}_{mn}
     &=& -\frac{2}{\sqrt{\bm C}}\frac{\delta S_M}{\delta {\bm C}^{mn}}
\,.
\label{Gmn:Tmn}
\end{eqnarray}
This formulation of the theory guarantees the (contracted) Bianchi
identity, which in the eight dimensional form reads,
\begin{eqnarray}\nonumber
   \bm\nabla^{m} \bm G_{mn}
   =
   0.
\label{Bianchi:8d}
\end{eqnarray}
The proof is analogous to that in general relativity. As a
consequence, the stress energy must be covariantly conserved,
${\bm \nabla}^m{\bm T}_{mn}=0$, just as desired. Note that
imposing the reciprocity symmetry on the action~(\ref{eq:complex
action:full}) (off-shell) would result in an over-constrained
on-shell dynamics which would fail to satisfy the Bianchi
identity~(\ref{Bianchi:8d}). We consider that as unacceptable,
since that would imply nonconservation of the stress energy
tensor, implying that energy would leak from our four dimensional
space-time hypersurface into the energy-momentum directions.

The stress energy tensor corresponding to the scalar field
action~(\ref{eq:scalar action:fields}) is just,
\begin{equation}
  {\bm T}_{mn} =  \alpha (\partial_m\Phi)^\dagger
                    \cdot\partial_n\Phi
          + \beta(\partial_m\Psi)^T\cdot\partial_n\Psi
          +{\bm C}_{mn} {\cal L}
\,,
\label{Tmn:scalars}
\end{equation}
where we used
$\delta \sqrt{\bm C} = -\frac12  \sqrt{\bm C}{\bm C}_{mn}\delta {\bm C}^{mn}$.

When written in the four dimensional notation, Eqs.~(\ref{eq:hermitian eom})
reduce to,
\begin{subequations}
\begin{eqnarray}
  G_{\mu\nu} = R_{\mu\nu}  &=& 8\pi G_N T_{\mu\nu}
\label{eq:Hermitian Einstein's equations:1}
\\
  G_{\mu\bar\nu} + C_{\mu\bar\nu}\Lambda &=& 8\pi G_N T_{\mu\bar\nu}
 \label{eq:Hermitian Einstein's equations:2}
\end{eqnarray}
\end{subequations}
plus the corresponding Hermitian conjugate equations, where
$C_{\mu\nu} = 0$ and $G_{\mu\bar\nu} = R_{\mu\bar\nu}-
\frac12C_{\mu\bar\nu}R$. We also have,
$R_{\mu\nu}=R^\alpha_{\;\mu\alpha\nu}+R^{\bar\alpha}_{\;\mu\bar\alpha\nu}$
and $R_{\mu\bar\nu} = R^\alpha_{\;\mu\alpha\bar\nu}
                + R^{\bar\alpha}_{\;\mu\bar\alpha\bar\nu}$.
Note that the holomorphic equation~(\ref{eq:Hermitian Einstein's equations:1})
does not admit a cosmological term $\Lambda$. Indeed, $\Lambda$
is removed from~(\ref{eq:Hermitian Einstein's equations:1})
by the on-shell reciprocity symmetry.

\subsection{Metric Compatibility}
\noindent When working in the first order
formalism~\cite{MantzThesis:2007}, in addition to
Eqs.~(\ref{eq:hermitian eom}) one also obtains the metric
compatibility equations,
\begin{eqnarray}\label{eq:metric compatibility equations}
    {\bm \nabla_m} {\bm C_{nr}} = {\bm 0}.
\end{eqnarray}
We list two components in four dimensional notation
\begin{eqnarray}
    \nabla_\rho C_{\mu\bar{\nu}} = 0
\,,\qquad
    \nabla_\rho C_{\mu\nu} = 0.
\label{eq:Hermitian metric compatibility equations}
\end{eqnarray}
One can check that the connection coefficients (\ref{eq:Hermitian
connection coefficients z basis 4 d}) are consistent with the
metric compatibility condition (\ref{eq:Hermitian metric
compatibility equations}): since $C_{\mu\nu}= 0$ and
$\Gamma^{\bar{\rho}}_{\mu\nu} = 0$, we obtain that $\nabla_\rho
C_{\mu\nu} = 0.$ The mixed components\footnote{In order to show
that $\nabla_\rho C_{\bar{\mu}\nu} = 0$, one needs to realize that
$C^{\mu\bar{\epsilon}}C_{\nu\bar{\epsilon}} \equiv \beta^\mu_\nu
\neq \delta^\mu_\nu = C^{\bar{\epsilon}\mu}C_{\nu\bar{\epsilon}}$,
since the Hermitian metric is nonsymmetric (it is Hermitian). The
action of $\beta$ on the metric, $\beta^\alpha_\nu
C_{\bar{\mu}\alpha} = C_{\nu\bar{\mu}}$, can be derived by
inserting the identity in the Hermitian line element
(\ref{eq:Hermitian familiar form line element}).} of the metric
compatibility condition
\begin{eqnarray}\nonumber
    \nabla_\rho C_{\mu\bar{\nu}}
    =
    \partial_\rho C_{\mu\bar{\nu}}
    -
    \Gamma^{\epsilon}_{\rho\mu}C_{\epsilon\bar{\nu}}
    -
    \Gamma^{\bar{\epsilon}}_{\rho\bar{\nu}}C_{\bar{\epsilon}\mu}
\end{eqnarray}
vanish as well
\begin{eqnarray}\nonumber
    \partial_\rho C_{\mu\bar{\nu}}
    -
    \frac{1}{2}(\partial_{\rho} C_{\mu\bar{\nu}}
    + \partial_{\mu} C_{\rho\bar{\nu}})
    -
    \frac{1}{2}(\partial_{\rho} C_{\mu\bar{\nu}}
    - \partial_{\mu} C_{\rho\bar{\nu}})
    =
    0.
\end{eqnarray}
On the other hand, writing the Hermitian metric component in terms
of vielbeins and using the Leibnitz rule for the covariant
derivative, one can show that the connection coefficients
(\ref{eq:Hermitian connection coefficients z basis 4 d}) do not
imply the vielbein compatibility (\ref{eq:vielbein compat})
discussed above. Indeed, writing Eq.~(\ref{eq:Hermitian metric
compatibility equations}) as
\begin{eqnarray}\nonumber
    \nabla_\rho (e_{\mu}) e_{\bar{\nu}}
    +
    e_{\mu}\nabla_\rho  e_{\bar{\nu}}
    =
    0,
\end{eqnarray}
implies that $
    \nabla_\rho  e_{\bar{\nu}}
    =
    -e_{\bar{\epsilon}}\Gamma^{\bar{\epsilon}}_{\rho\bar{\nu}}
    - e_{\epsilon}\Gamma^{\epsilon}_{\rho\bar{\nu}}
    \neq 0
$ is not zero for instance.

\section{Counting Degrees of freedom}
\noindent In order to get a glimpse of the general structure of
Hermitian gravity, it might be educational to determine some
properties of relevant objects, which are part of the theory. If
we would like to know for instance, if we can always go to a
freely falling frame, we can begin with counting the degrees of
freedom of an arbitrary coordinate transformation of the metric
tensor, in order to see if there are enough coordinate degrees of
freedom in order to do so. If we then Taylor expand both sides of
the coordinate transformation of the metric tensor
\begin{eqnarray}\label{eq:coordinate trans count}
    {\tilde{C}}_{\bar{\mu}\nu}
    =
    \frac{\partial z^{\bar{\alpha}}}{\partial {\tilde{z}}^{\bar{\mu}}}
    \frac{\partial z^{\beta}}{\partial {\tilde{z}}^{\nu}}
    C_{\bar{\alpha}\beta},
\end{eqnarray}
we can collect terms of a specific order of the expansion of both
sides of the equation and equate these terms. We are Taylor
expanding both sides of the coordinate transformation of the
Hermitian metric (\ref{eq:coordinate trans count}) around a point
$p$ on the (smooth) manifold.

When considering the zeroth order terms of the expansion,
\begin{eqnarray}\label{eq:coordinate trans count}
    {\tilde{C}}_{\bar{\mu}\nu} |_p
    =
    \frac{\partial z^{\bar{\alpha}}}{\partial {\tilde{z}}^{\bar{\mu}}}
    \frac{\partial z^{\beta}}{\partial {\tilde{z}}^{\nu}}
    C_{\bar{\alpha}\beta}\Big |_p,
\end{eqnarray}
we have $16$ degrees of freedom at the left hand side of the
equation, since the metric is Hermitian. The formula for the real
degrees of freedom of a Hermitian matrix is $d^2$, where $d$ is
the complex dimension of the manifold, which is 4 in this case. On
the right hand side of the equation, there are 32 real degrees of
freedom to transform to the flat space metric (there are $32$
degrees of freedom instead of 64, because the coordinate
transformations are holomorphic, implying they satisfy the
Cauchy-Riemann equations). Subtracting the two, we obtain $32-16 =
16$ degrees of freedom, which leave the flat space metric
invariant. These 16 degrees of freedom are precisely the 16
degrees of freedom of the $U(1,3)$ group, which by definition
leave the Hermitian flat space metric invariant.

Considering the following terms at first order of the expansion of
the coordinate transformation (\ref{eq:coordinate trans count})
\begin{eqnarray}\nonumber
    \bm {\tilde{\partial}}_e \bm {\tilde{C}}_{m n}|_p
    + \cdots
    =
    \frac{\bm \partial^2 \bm z^{a}}{\bm \partial \bm {\tilde{z}}^{e} \bm \partial \bm {\tilde{z}}^{m}}
    \frac{\bm \partial \bm z^{b}}{\bm \partial \bm {\tilde{z}}^{n}}
    \bm C_{a b}\Big |_p
    +\cdots,
\end{eqnarray}
we count $64$ real degrees of freedom on the left hand side and
$80$ on the right. We obtain the $64$ real dimension in the
following manner. The complex number of degrees of freedom for a
Hermitian matrix is $\frac{1}{2}d^2$, where $d$ is again the
complex dimension of the manifold, which is, as said before, $4$
in this case. The complex dimension of the partial derivative is
$d$. When multiplying these numbers we obtain $32$ complex degrees
of freedom, which is equivalent to $64$ real degrees of freedom.
The $80$ degrees of freedom from the first factor of the term on
the right hand side are obtained as follows. The numerator has $d$
complex degrees of freedom. The denominator has
$\frac{1}{2}d(d+1)$ complex degrees of freedom, which is just the
formula of a symmetric matrix, since partial derivatives commute.
By multiplying these numbers together we obtain $4\cdot 10 = 40$,
complex degrees of freedom, which is equivalent to $80$ real
degrees of freedom. When subtracting these numbers we obtain $80 -
64 = 16$ real degrees of freedom. This means that we have $16$
degrees of freedom too many in order to transform to the free
falling frame.

Finally, considering the following two terms at second order of
the expansion of the coordinate transformation of the Hermitian
metric (\ref{eq:coordinate trans count})
\begin{eqnarray}\nonumber
    \bm {\tilde{\partial}}_e \bm {\tilde{\partial}}_f\bm {\tilde{C}}_{m n} |_p
    + \cdots
    =
    \frac{\bm \partial^3 \bm z^{a}}{\bm \partial \bm {\tilde{z}}^{e}
    \bm \partial \bm {\tilde{z}}^{f} \bm \partial \bm {\tilde{z}}^{m}}
    \frac{\bm \partial \bm z^{b}}{\bm \partial \bm {\tilde{z}}^{n}}
    \bm C_{a b}\Big|_p
    +\cdots,
\end{eqnarray}
we have $160$ real degrees of freedom at the left hand side and
$160$ on the right. We obtain the $160$ real degrees of freedom on
the left hand side as follows. The complex degrees of freedom of
the partial derivatives is $ \frac{1}{2}d(d+1) $ and the complex
degrees of freedom of the metric is again $\frac{1}{2}d^2$.
Multiplying these numbers together we obtain $10 * 8 = 80$ complex
degrees of freedom, which is equivalent to $160$ real degrees of
freedom. The $160$ real degrees of freedom on the right hand side
are obtained as follows. The numerator has again dimension $d$.
The denominator has dimension $\frac{1}{3!}d(d+1)(d+2)$.
Multiplying these numbers together we obtain $4*20 = 80$ complex
degrees of freedom, which is again equivalent to $160$ real
degrees of freedom. When subtracting these two numbers we obtain
$160 - 160 = 0$ degrees of freedom. This means that we have
precisely enough degrees of freedom in order to obtain flat space
at second order of the expansion. Hence there is no
space-time-momentum-energy curvature in the theory of Hermitian
gravity. This might appear as problematic since general relativity
does contain space-time curvature, which we cannot get rid off by
coordinate transformations. We shall see below that in the limit
of projecting space-time-momentum-energy onto space-time we will
obtain space-time curvature as an artifact of the limiting
procedure.

Hermitian gravity is dynamical in the sense that there are second
order derivatives, acting upon the dynamical variable, the
vielbein or the metric. Consider the following independent
components of the Hermitian Riemann tensor,
\begin{eqnarray}\label{eq:Riemann un mixed indices}
    R^{\kappa}_{\lambda\mu\nu}
    =
    C^{\bar{\epsilon}\kappa}\partial_{\lambda }\partial_{[\mu}C_{\nu]\bar{\epsilon}}
    +
    \text{first order derivatives},
\end{eqnarray}
and its Hermitian conjugate. Unlike the components of the Riemann
tensor (\ref{eq:Riemann known}), constructed from the known
complex connection coefficients (\ref{eq:connection known}), we do
have non-vanishing components of the Riemann tensor, which do
contain second order derivatives. These components \textit{do}
enter the Hermitian Einstein equations (\ref{eq:Hermitian
Einstein's equations:2}), although $C_{\mu\nu} = 0 $. These
components of the Hermitian Einstein tensor act as constraints,
such that we remain on the hypersurface, which specified by the
reciprocity transformation.

\section{The Limit to General Relativity}
\noindent The limit of Hermitian gravity to the theory of general
relativity is based on the assumption that the $y$ coordinate and
its corresponding vielbein are small. When expanding these
theories in powers of $y$ and its corresponding vielbein, we would
hope to obtain the theory of general relativity at zeroth order of
the expansion and meaningful corrections to the theory at linear
order. We will see that this is not the case, since we will obtain
corrections to general relativity at zeroth order. The easiest way
to obtain the limit to general relativity is to expand the real
and the imaginary parts of the vielbein in terms of the $y$
coordinate in order to collect the terms in orders of the $y$
coordinate and its corresponding vielbein, ${e_I}_{\check{\mu}}$,
yielding
\begin{subequations}\label{eq:vielbein expansion}
\begin{eqnarray}\label{eq:expansion x vielbein}
    {e_{R}}_\mu (x,y) &=& {e_{R}}_\mu (x) - y^\lambda
    \partial_\lambda {e_{I}}_{\check{\mu}} + O(y^2)
    \\\nonumber
    &=& {e_{R}}_\mu (x) + O'(y^2)
\end{eqnarray}
and
\begin{eqnarray}\label{eq:expansion y vielbein}
    {e_{I}}_{\check{\mu}} (x,y) = {e_{I}}_{\check{\mu}} (x) + y^\lambda
    \partial_\lambda {e_{R}}_\mu (x)
   + O(y^2).
\end{eqnarray}
\end{subequations}
These expansions contain sufficient information in order to obtain
the limit to general relativity. We will, however, also expand the
rotated metric components and a component of the rotated
connection coefficients. Using the expansions of the real and
imaginary parts of the vielbeins (\ref{eq:vielbein expansion}),
the rotated metric components up to second order of the $y$
coordinate and its corresponding vielbein are
\begin{eqnarray}\nonumber
    g_{\mu\nu}(x,y)
    =
    g_{\mu\nu}(x) + O(y^2),
\end{eqnarray}
\begin{eqnarray}\nonumber
    g_{\mu\check{\nu}}(x,y)
    =
    g_{\mu\check{\nu}}(x) + y^{\lambda}\left(\partial_{\lambda}(e_{\mu})e_{\nu} - e_{\mu}\partial_{\lambda}e_{\nu} \right)  + O(y^2),
\end{eqnarray}
\begin{eqnarray}\nonumber
    g_{\check{\mu}\nu}(x,y)
    =
    g_{\check{\mu}\nu}(x)  + y^{\lambda}\left(e_{\mu}\partial_{\lambda}e_{\nu}- \partial_{\lambda}(e_{\mu})e_{\nu}\right) + O(y^2),
\end{eqnarray}
and
\begin{eqnarray}\nonumber
    g_{\check{\mu}\check{\nu}}(x,y)
    =
     g_{\check{\mu}\check{\nu}}(x) + O(y^2),
\end{eqnarray}
where $
    g_{\mu\nu}(x)
    =
    g_{\check{\mu}\check{\nu}}(x)
    =
    e_{\mu}e_{\nu}(x)
$ and where $
    g_{\check{\mu}\nu}(x)
$ and $
    g_{\mu\check{\nu}}(x)
$ can be just read off the expression of the rotated Hermitian
metric in terms of vielbeins. Using again the expansions of the
real and imaginary parts of the vielbeins (\ref{eq:vielbein
expansion}), the  $\Gamma ^\rho_{\mu\nu} (x,y)$ component of the
connection coefficients (\ref{eq:Hermitian connection coefficients
z basis 4 d}) up to second order of the $y$ coordinate and its
corresponding vielbein is
\begin{eqnarray}\nonumber
   \Gamma ^\rho_{\mu\nu} (x,y) =  \Gamma ^\rho_{\mu\nu} (x) +
   O(y^2),
\end{eqnarray}
where $\Gamma ^\rho_{\mu\nu} (x)$ is just the ordinary
Levi-Civit\`a connection. With the expansions of the connection
coefficients, we can now check if the theory of Hermitian gravity
reduces to the theory of general relativity by plugging them into
the rotated Hermitian geodesic equation, keeping only terms of
linear order in the $y$ coordinate and its corresponding vielbein,
yielding the ordinary geodesic equation
\begin{eqnarray}\nonumber
    \ddot{x}^\rho + \Gamma_{\mu\nu}^\rho(x){\dot{x}}^\mu {\dot{x}}^\nu
    + O(y^2) = 0
\end{eqnarray}
without any first order corrections present. Though the result is
not spectacular at first sight, it should be pleasing that the
theory of Hermitian gravity reduces to the well tested theory of
general relativity, for the Hermitian geodesic equation. In order
to see if the theory predicts any interesting new physics we have
to collect terms up to second order, yielding
\begin{eqnarray}\nonumber
    \ddot{x}^\rho &+& \Gamma_{\mu\nu}^\rho(x,y){\dot{x}}^\mu
    {\dot{x}}^\nu + [\Gamma^\rho_{\check{\mu}\nu} (x,y)+
    \Gamma^\rho_{\nu\check{\mu}}(x,y)]\dot{y}^{\check{\mu}}\dot{x}^\nu
    \\\nonumber
    &+&
    \Gamma^\rho_{\check{\mu}\check{\nu}}(x,y) \dot{y}^{\check{\mu}}\dot{y}^{\check{\nu}}
    +
    O(y^3)
    =
    0,
\end{eqnarray}
where the connection coefficients $\Gamma^\rho_{\check{\mu}\nu}
(x,y),$ $ \Gamma^\rho_{\nu\check{\mu}}(x,y)$ and $
\Gamma^\rho_{\check{\mu}\check{\nu}}$ are just the connection
coefficients expanded up to linear order~\cite{MantzThesis:2007}, but
where the connection coefficient $\Gamma_{\mu\nu}^\rho(x,y)$
has to be expanded up to second order since the term
${\dot{x}}^\mu {\dot{x}}^\nu$, multiplying
$\Gamma_{\mu\nu}^\rho(x,y)$, is of zeroth order in the $y$
coordinate and its corresponding vielbein.

The Hermitian Einstein's equations get corrections to the
Einstein's equations of general relativity at zeroth order. This
can be seen when considering the rotated Hermitian Ricci tensor
\begin{eqnarray}\nonumber
    R_{\mu\nu}
    =
    R^{\lambda}_{\phantom{{\lambda}}\mu\lambda\nu} + R^{\check{\lambda}}_{\phantom{{\lambda}}\mu\check{\lambda}\nu}
    =
    R^{GR}_{\mu\nu} +
    {R_{\text{cor}}}^{\check{\lambda}}_{\phantom{{\lambda}}\mu\check{\lambda}\nu} +
    O(y^2),
\end{eqnarray}
where $R^{GR}$ is the Ricci tensor according to general relativity
and where $R_{\text{cor}}$ are the terms of
$R^{\bar{\lambda}}_{\phantom{{\lambda}}\mu\bar{\lambda}\nu}$ of
zeroth order in the expansion in the $y^{\check{\mu}}$ coordinate.
Similarly the rotated Hermitian Einstein tensor
\begin{eqnarray}\nonumber
    G_{\mu\nu}
    =
    G^{GR}_{\mu\nu} +
    {R_{\text{cor}}}^{\check{\lambda}}_{\phantom{{\lambda}}\mu\check{\lambda}\nu}
    -\frac{1}{2}g_{\mu\nu}g^{\alpha\beta}{R_{\text{cor}}}^{\check{\lambda}}_{\phantom{{\lambda}}\alpha\check{\lambda}\beta}
    \\\nonumber
    -
    \frac{1}{2}g_{\mu\nu}g^{\check{\alpha}\check{\beta}}
    \big({R_{\text{cor}}}^{\lambda}_{\phantom{{\lambda}}\check{\alpha}\lambda\check{\beta}}
    +
    {R_{\text{cor}}}^{\check{\lambda}}_{\phantom{{\lambda}}\check{\alpha}\check{\lambda}\check{\beta}}\big)]
    +
    O(y^2),
\end{eqnarray}
gets corrections of zeroth order in the expansion in the
$y^{\check{\mu}}$ coordinate. Hence, at this point one needs to
look at the solutions of Hermitian gravity in order to see if the
theory contradicts experiment or not.

\section{Hermitian Cosmology}
\label{Hermitian Cosmology}

\noindent
In order to describe our Universe correctly,
which is isotropic and homogeneous on large scales,
our complex theory should permit solutions that
possess the symmetries of isotropy and homogeneity
and furthermore these solutions should correctly reduce to
the Friedmann-Lema{\^\i}tre-Robertson-Walker
(FLRW) cosmology at low energies.
Since the vielbeins of our theory are holomorphic functions,
we can demand spatial isotropy and homogeneity
by implementing a scale factor that is a
holomorphic function of the time-like coordinate only,
\begin{eqnarray}
    e^{a}_{\mu}
    =
    a(z^0)\delta_{\mu}^{a}
    \phantom{halloda}
    e^{a}_{\bar{\mu}}
    =
    a(z^{\bar{0}})\delta_{\bar{\mu}}^{a}
\,,
\label{eq:vielbein:cosmology}
\end{eqnarray}
where the complex scale factor can be specified in terms of its
real and imaginary parts
\begin{eqnarray}\nonumber
    a(z^0)
    =
    a_{R}(z^0)
    +
    i a_{I}(z^0)
    \phantom{hallo}
    \bar{a}(z^{\bar{0}})
    =
    a_{R}(z^{\bar{0}})
    -
    i a_{I}(z^{\bar{0}}).
\end{eqnarray}
Note that $z^0 = t + i\frac{G_N}{c^4} E $. The {\it
Ansatz}~(\ref{eq:vielbein:cosmology}) yields a cosmology with flat
spatial sections, which suffices for our purpose.\footnote{ To
generalist the {\it Ansatz}~(\ref{eq:vielbein:cosmology}) to
space-times with a constant spatial curvature, one would have to
replace $a(z^0)\delta_\mu^a$ in Eq.~(\ref{eq:vielbein:cosmology})
by the corresponding vielbein whose spatial indices describe the
geometry of a static 3-sphere (3-hyperboloid) for a space with
positively (negatively) curved spatial sections. Thus for a
space-time with positively curved spatial sections ($\kappa>0$) we
have, $e_\mu = a(z^0)[\delta_\mu^{0}+\delta_\mu^{\chi}
         +(1/\sqrt{\kappa})\sin(\sqrt{\kappa}\chi)\delta_\mu^{\theta}
         +(1/\sqrt{\kappa})\sin(\sqrt{\kappa}\chi)\sin(\theta)\delta_\mu^{\varphi}]$,
where $\chi\in[0,\pi/\sqrt{\kappa}]$, $\theta\in[0,\pi]$ and
$\varphi \in[0,2\pi)$ are the spherical coordinates on $S^3$. For
a space with a negative curvature ($\kappa<0$) the tetrad $e_\mu$
is obtained from the tetrad of the closed universe with the
replacement, $(1/\sqrt{\kappa})\sin(\sqrt{\kappa}\chi)\rightarrow
(1/\sqrt{-\kappa})\sinh(\sqrt{-\kappa}\chi)$, where now
$\chi\in[0,\infty)$.} With this {\it Ansatz} for the vielbein,
 the connection coefficients become holomorphic (or anti-holomorphic)
functions. Consider for example the following two connection coefficients
\begin{eqnarray}\label{eq:connection H Cosmo unrotated}
    \Gamma^{\rho}_{\mu\nu}
    &=&
    \frac{1}{2}\frac{a'}{a}(\delta^{\rho}_{\nu}\delta^{0}_{\mu}
    +
    \delta^{0}_{\nu}\delta^{\rho}_\mu)
    \\\nonumber
    \Gamma^{\rho}_{\bar{\mu}\nu}
    &=&
    \frac{1}{2}\frac{\bar{a}'}{\bar{a}}(\delta^{\bar{0}}_{\bar{\mu}}\delta^{\rho}_{\nu}
    -
    \eta^{\bar{0}\rho}\eta_{\bar{\mu}\nu}).
\end{eqnarray}
These expressions for the connection coefficients can then be used
to obtain the components of the Hermitian Ricci tensor. The
expression for the mixed components of the Ricci tensor is then
\begin{eqnarray}\nonumber
   R_{\bar{\mu}\nu}
   =
   \frac{\bar{a}'a'}{\bar{a}a}
   \left[\left(\frac{ d - 1}{2}\right)\delta^{\bar{0}}_{\bar{\mu}}\delta^{0}_{\nu} + (d-1)\eta_{\bar{\mu}\nu}\right],
\end{eqnarray}
where $d$ is again the complex dimension of the manifold. Taking
$d$ to be four, we obtain the following expressions for the
independent mixed components of the Hermitian Ricci
tensor\footnote{Here the Latin indices $i$ and $j$ take values
$1,2,3$.}
\begin{eqnarray}\nonumber
    R_{\bar{0}0}
    =
    -\frac{3}{2}\frac{\bar{a}'a'}{\bar{a}a}
    \phantom{halloda}
   R_{\bar{i}j}
   =
   3\frac{\bar{a}'a'}{\bar{a}a} \eta_{\bar{i}j}
\end{eqnarray}
and the following expression for the Hermitian Ricci scalar
\begin{eqnarray}\nonumber
    R
    =
    C^{\bar{\mu}\nu}R_{\bar{\mu}\nu}
    +
    C^{\mu\bar{\nu}}R_{\mu\bar{\nu}}
    =
    21 \frac{\bar{a}'a'}{(\bar{a} a)^2}.
\end{eqnarray}
The nonzero unmixed components are then
\begin{eqnarray}\nonumber
   R_{00}(z^0)
   =
   \frac{9}{2}\left(\frac{a'}{a}\right)^2 - 3\left(\frac{a''}{a}\right)
\end{eqnarray}
and its complex conjugate. The independent components of the
Hermitian Einstein tensor then become
\begin{eqnarray}\nonumber
    &&G_{\bar{0}0}
    =
    9\frac{\bar{a}'a'}{\bar{a}a}
\,,
    \phantom{halloda}
    G_{\bar{i}j}
    =
    -\frac{15}{2}\frac{\bar{a}'a'}{\bar{a}a}\eta_{\bar{i}j}
    \\\label{eq:einstein t H cosmo unrotated}
    &&G_{00}
     =
    \frac{9}{2}\left(\frac{a'}{a}\right)^2 - 3\left(\frac{a''}{a}\right)
\,
\end{eqnarray}
and their complex conjugates.
If we now make use of the connection
coefficients~(\ref{eq:connection H Cosmo unrotated})
we can easily check that
the Einstein tensor of Hermitian
gravity~(\ref{eq:einstein t H cosmo unrotated}) obeys
the Bianchi identity~(\ref{Bianchi:8d}), implying that
the Einstein tensor is divergenceless, as it should be.
This represents a nontrivial check of the accuracy of our calculation.

 Consider now the matter action. For definiteness (and simplicity) we shall
consider the two scalar field action~(\ref{eq:scalar action}).
Since in standard general relativity one can obtain
any desired expansion $a=a(t)$ by appropriately choosing the
scalar field potential,
we expect that the action~(\ref{eq:scalar action}) does not pose
any important restrictions to the Hermitian cosmology.

  The underlying symmetries of a (flat) FLRW cosmology
together with holomorphy then imply that the scalar fields are of
the form, $\phi=\phi(z^0)$ and $\psi=\psi(z^0)$. With this
observation we get that the nonvanishing components of the stress
energy tensor~(\ref{Tmn:scalars}) are,
\begin{eqnarray}
  T_{\mu\bar\nu} &=&  \alpha \delta_\mu^0\delta_{\bar\mu}^{\,\bar 0}
                         \phi' \bar\phi'
              + \eta_{\mu\bar\nu}\big(\alpha\phi' \bar\phi'- a\bar a V\big)
\nonumber
\\
  T_{\mu\nu} &=&  \beta \delta_\mu^0\delta_{\mu}^{\,0}
                         {\psi'}^2
\,,
\label{Tmn:scalars:cosmology:2}
\end{eqnarray}
plus the Hermitian conjugates. Here we used
$\phi'=(\partial/\partial z^0)\phi$ and
$\bar\phi'=(\partial/\partial z^{\bar 0})\bar\phi$. The
nonvanishing components in~(\ref{Tmn:scalars:cosmology:2}) are,
\begin{eqnarray}\label{Tmn:scalars:cosmology:3}
  T_{0\bar 0} = a\bar a V
\phantom{ha}
  T_{i\bar j} = \delta_{i\bar j}\big(\alpha\phi' \bar\phi'- a\bar a V\big)
\phantom{ha}
  T_{00} =  \beta{\psi'}^2
.
\end{eqnarray}
 When combined with Eqs.~(\ref{eq:einstein t H cosmo unrotated})
these yield the following equations for Hermitian
cosmology~(\ref{eq:Hermitian Einstein's equations:1}--\ref{eq:Hermitian Einstein's equations:2}),
\begin{subequations}
\begin{align}
  &G_{00} \equiv
\frac{9}{2}\left(\frac{a'}{a}\right)^2 - 3\left(\frac{a''}{a}\right)
  = 8\pi G_N \beta {\psi'}^2
\label{eq:Hermitian Einstein's equations:3}
\\
  &G_{0\bar0}
  +
  C_{0\bar0}\Lambda
  \equiv
  9\frac{\bar{a}'a'}{\bar{a}a} - a\bar a \Lambda
  =
  8\pi G_N a\bar a V
\label{eq:Hermitian Einstein's equations:4}
\\\nonumber
  &\frac13\delta^{i\bar j}\big(G_{i\bar j} + C_{i\bar
  j}\Lambda\big)
    \equiv
    -\frac{15}{2}\frac{\bar{a}'a'}{\bar{a}a}
   +
   a\bar a\Lambda
\\
    &=
    8\pi G_N \big(\alpha\phi' \bar\phi'- a\bar a V\big) \,,
\label{eq:Hermitian Einstein's equations:5}
\end{align}
\end{subequations}
which together with the scalar field equations of motion,
\begin{eqnarray}
  -3\alpha\frac{\bar a'}{a\bar a^2}\phi' - \partial_{\bar\phi}V &=& 0
\nonumber\\
  -3\beta\frac{\bar a'}{a\bar a^2}\psi' - \partial_{\psi}V &=& 0
\label{eq:scalar eom:cosmology}
\end{eqnarray}
represent the closed system of equations of Hermitian cosmology
with scalar fields. These equations are obtained by varying the
matter action~(\ref{eq:scalar action}) with respect to $\bar\phi$
and $\psi$, respectively. The scalar equations of
motion~(\ref{eq:scalar eom:cosmology}) can be can be also obtained
from the covariant stress-energy conservation. Inspired by the
form of stress-energy in FLRW spaces,
$T_{\mu\nu}=a^2\delta_\mu^{\;0}\delta_\nu^{\;0}(\rho+p)+p
g_{\mu\nu}$, the appropriate Hermitian gravity generalization is
of the form,
\begin{eqnarray}\nonumber
  T_{\mu\nu}
  &=&
  (\rho_h+p_h)a^2\delta_\mu^{\;0}\delta_\nu^{\;0}
  \\
  T_{\mu\bar\nu}
  &=&
  a^2\delta_\mu^{\;0}\delta_{\bar\nu}^{\;\bar 0}(\rho+p)
   +p C_{\mu\bar\nu}
   \,.
\label{stress energy:Hermitian}
\end{eqnarray}
Comparing this with Eqs.~(\ref{Tmn:scalars:cosmology:3}) then implies
\begin{eqnarray}\label{Tmn:scalars:cosmology:p+rho}
  \rho = V
\phantom{hall}
  p = \alpha\dot\phi\dot{\bar\phi}- V
\phantom{hall}
  \rho_h+p_h =  \beta{\dot\psi}^2
\end{eqnarray}
where $(1/a)\phi' = \dot \phi$ and $(1/a)\psi' = \dot \psi$.
While the pressure has a standard form, note that the kinetic term does not
contribute to the energy density of Hermitian gravity.~\footnote{
One would arrive at a more standard expression for the scalar energy
density if one would replace $\rho\rightarrow (\rho-p)/2$
in Eq.~(\ref{stress energy:Hermitian}).}
The stress energy conservation $\nabla^m T_{mn}=0$ then implies,
\begin{equation}
 \dot\rho + 3 \bar H (\rho_h+p_h) + 3 H(\rho+p) = 0
\end{equation}
and the complex conjugate equation, where
$\dot{\rho}=(1/a)\partial_{0}\rho$ and where $H=a'/a^2 = \dot a/a$
and $\bar H = \bar a'/{\bar a}^2 = \dot {\bar a}/\bar a$. This
together with Eqs.~(\ref{Tmn:scalars:cosmology:p+rho}) implies the
scalar field equations~(\ref{eq:scalar eom:cosmology}), which
checks the consistency of our formulation of Hermitian gravity
with scalar fields.

Next, it is convenient to divide equation~(\ref{eq:Hermitian
Einstein's equations:3})
 by $a^2$ and
Eqs.~(\ref{eq:Hermitian Einstein's equations:4}--\ref{eq:Hermitian Einstein's equations:5})
by $a\bar a$, respectively. Combining Eqs.~(\ref{eq:Hermitian Einstein's equations:4})
and~(\ref{eq:Hermitian Einstein's equations:5}) results in
the constraint equations,
\begin{subequations}
\begin{eqnarray}
  H\bar H &=& \frac{8\pi G_N}{9}\big(V+\lambda\big)
\,,\qquad \lambda = \frac{\Lambda}{8\pi G_N}
\label{eq:herm:cosmology:1}
\\
  \dot\phi\dot{\bar\phi} &=& \frac{1}{6\alpha}\big(V+\lambda\big)
\,. \label{eq:herm:cosmology:2}
\end{eqnarray}
Equation~(\ref{eq:Hermitian Einstein's equations:3}) can be recast
as,
\begin{eqnarray}
-\dot H -\frac12 H^2 &=& \frac{8\pi G_N}{3} \beta {\dot \psi}^2
\,.
\label{eq:herm:cosmology:3}
\end{eqnarray}
With a help of Eq.~(\ref{eq:herm:cosmology:1}), the scalar
equations~(\ref{eq:scalar eom:cosmology})
become,

\begin{eqnarray}
 \dot\phi &=& - \frac{3H}{8\pi G_N\alpha}
           \frac{\partial_{\bar\phi}V}{V+\lambda}
\label{eq:herm:cosmology:4}
\\
 \dot\psi &=& - \frac{3H}{8\pi G_N\beta}\frac{\partial_{\psi}V}{V+\lambda}
\,.
\label{eq:herm:cosmology:5}
\end{eqnarray}
\end{subequations}
Equations~(\ref{eq:herm:cosmology:1}--\ref{eq:herm:cosmology:5})
(and their hermitean conjugates) are the
fundamental equations of hermitean cosmology. Note that
there are 5 (2 real and 3 complex) equations for 3 complex quantities
$H$, $\phi$ and $\psi$, so the system is overdetermined, and there is
no guarantee that a solution exists.

 We shall now show that a solution exists, and moreover we shall explicitly
construct a class of solutions that gives rise to a power law expansion
of the scale factor $a=a(z^0)$.

 Firstly, Eq.~(\ref{eq:herm:cosmology:1}) implies that
$\bar V = V$ and $\bar\lambda=\lambda$ are real. Secondly,
Eq.~(\ref{eq:herm:cosmology:4}) and its complex conjugate
imply that $\partial_{\bar \phi} \ln\big(V+\lambda\big)$
is a holomorphic functions of $\phi$ and that
$\partial_{\phi} \ln\big(V+\lambda\big)$ is an
antiholomorphic function of $\bar\phi$.
Consequently the potential is determined to be of the form,
\begin{equation}
 \ln(V+\lambda) = A_1 \phi + \bar A_2 \bar \phi + A_3
\,,
\end{equation}
where $A_1$, $A_2$ and $A_3$ are complex constants independent on
$\phi$ (still possibly dependent on $\psi$ and $\bar\psi$). The
reality of $V$ and $\lambda$ then implies that $A_1=A_2\equiv
\Omega$. Writing $A_3$ as $A_3=\ln(W)$ we have,
\begin{equation}
  V = -\lambda + W\exp\big[\Omega\big(\phi+\bar\phi\big)\big]
\,.
\label{eq:V:1}
\end{equation}
Finally, since $V$ is real, $W$ and $\Omega$ must be real
functions of $\psi$ and $\bar\psi$. Holomorphy is such a powerful
symmetry
 that - even though Eq.~(\ref{eq:V:1}) represents the most general solution
to Eqs.~(\ref{eq:herm:cosmology:4}) and its complex conjugate
-- the potentials $V$ for $\phi$ is, up to two `constants' $W$ and $\Omega$,
completely fixed.

 Now multiplying Eq.~(\ref{eq:herm:cosmology:4}) with its complex
conjugate, making use of Eq.~(\ref{eq:herm:cosmology:1}) and inserting
the resulting equation into~(\ref{eq:herm:cosmology:2}),
we obtain,
\begin{equation}
  \frac{4\pi G_N\alpha}{3} = \partial_{\bar \phi}\ln\big(V+\lambda)
                        \partial_{\phi}\ln\big(V+\lambda)
                     \equiv \Omega^2
\,,
\label{eq:omega}
\end{equation}
implying that there are two allowed values for $\Omega$,
\begin{equation}
\Omega_\pm \equiv \pm \omega = \pm \sqrt{\frac{4\pi G_N\alpha}{3}}
\,,
\label{eq:omega2}
\end{equation}
fixing thus $\Omega$ completely (up to a sign).
(This sign ambiguity reflects the symmetry
of the theory~(\ref{eq:scalar action})
under the transformation,
$\phi\rightarrow -\phi$, $\bar \phi \rightarrow -\bar\phi$.)
With this
Eq.~(\ref{eq:herm:cosmology:4}) becomes
\begin{equation}
  \dot \phi = \mp\frac{1}{2\omega}\frac{d}{dt}\ln(a)
\,,
\nonumber
\end{equation}
which is solved by
\begin{equation}
   a=a_0 {\rm e}^{\mp 2\omega (\phi-\phi_0)}
   \,.
\label{avsphi}
\end{equation}
($\phi_0$ is unphysical as it can be absorbed in the definition of
$a_0$.) This means that $\phi$ is not an independent field, but a
constrained field which is just a reparametrization of the scale
factor $a$. This is not surprising, given the fact that $\phi$
solves the constraint equations of Hermitian gravity.

\medskip

The remaining equations to be solved
are~(\ref{eq:herm:cosmology:3}) and~(\ref{eq:herm:cosmology:5}),
and the remaining freedom in the potential is in
$W=W(\psi,\bar\psi)$. Similarly as above, we can see from
Eq.~(\ref{eq:herm:cosmology:5}) that $\partial_\psi \ln(V+\lambda)
=
\partial_\psi\ln(W) $ must be a holomorphic function of $\psi$. We
conclude that $W$ must be a product of a holomorphic function of
$\psi$ and an antiholomorphic function of $\bar\psi$ (and they
must be mutually equal),
\begin{equation}
  W = w(\psi)\bar w(\bar\psi)
\,.
\label{Eq:W}
\end{equation}
With this observation, making use of~(\ref{eq:herm:cosmology:5}),
 Eq.~(\ref{eq:herm:cosmology:3}) can be recast as,
\begin{equation}
\epsilon \equiv -\frac{\dot H}{H^2} = \frac12
       + \frac{3}{8\pi G_N\beta}\Big(\partial_\psi\ln(w(\psi))\Big)^2
\,.
\label{eq:herm:cosmology:3b}
\end{equation}
Let us now consider the case when $\epsilon = {\rm const.}$
In this case Eq.~(\ref{eq:herm:cosmology:3b}) can be easily
solved for $w(\psi)$ in terms of $\epsilon$,
\begin{equation}
 w
  = w_0\exp\bigg[\sqrt{\frac{8\pi G_N\beta}{3}\Big(\epsilon-\frac12\Big)}\,\psi
           \bigg]
           \,,
\label{eq:w:power-law}
\end{equation}
where $w_0$ is a field independent constant.
The potential $V$ that yields a power law
expansion is therefore given by
\begin{eqnarray}\label{eq:V:power-law}
    V
    &=&
    - \frac{\Lambda}{8\pi G_N}
    +
    V_0 \exp\big[\pm\omega(\phi+\bar\phi)\big]\times
    \\\nonumber
    &\times&
    \exp\left[\sqrt{\frac{8\pi G_N\beta}{3}}
                 \bigg(\Big(\epsilon-\frac12\Big)^{1/2}\,\psi
                    +  \Big(\bar\epsilon-\frac12\Big)^{1/2}\,\bar\psi
                 \bigg)\right]
\,,
\end{eqnarray}
where $V_0=w_0\bar w_0$ is a (real) constant and $\omega =
\sqrt{4\pi G_N \alpha/3}$. Note that nonvanishing
$\text{Im}[\epsilon]$ breaks the charge-parity (CP) symmetry of
the $\psi$ field, as can be seen from the potential
(\ref{eq:V:power-law}). However, CPT is conserved. We will discuss
below how $\text{Im}[\epsilon]$ breaks time reversal symmetry, T.
Recall that a power law expansion means
\begin{equation}
  \epsilon \equiv \frac{3}{2}(1+w_f) = {\rm const.}
\,,
\label{eq:epsilon=const}
\end{equation}
where $w_f=p/\rho$ is the (complex) equation of state parameter of a
`cosmological fluid'
with a `pressure' $p$ and an `energy density' $\rho$.
Since $\epsilon = (d/dt)(1/H)$, a constant epsilon implies
a power law expansion with,
\begin{equation}
 H = \frac{1}{\epsilon z^0}
\,,\qquad
 a(z^0) = a_0\, \Big(\frac{z^0}{\zeta_0}\Big)^{1/\epsilon}
\,,
\label{eq:power law expansion}
\end{equation}
where $a_0$ and $\zeta_0$ are (complex) constants. In standard cosmology,
$\epsilon=3/2$ ($\epsilon=2$) correspond to matter (radiation) era,
while $0<\epsilon\ll 1$ corresponds to a slow roll inflation.

 From Eqs.~(\ref{eq:herm:cosmology:4}--\ref{eq:herm:cosmology:5})
and~(\ref{eq:V:power-law}) we find,
\begin{equation}
 \dot\phi = \mp \frac{H}{2\omega}
\,,\qquad
 \dot\psi = - \frac{H}{2\omega}\sqrt{\frac{\alpha}{\beta}(2\epsilon-1)}
\,,
\nonumber
\end{equation}
such that in a Universe expanding as a power law the two fields are not independent,
\begin{equation}
  \psi-\psi_0 = \pm \sqrt{\frac{\alpha}{\beta}(2\epsilon-1)}(\phi-\phi_0)
\,.
\label{eq:fields relation}
\end{equation}
When this is inserted into Eq.~(\ref{avsphi}) we immediately get
\begin{equation}
  a=a_0\exp\bigg[-\sqrt{\frac{16\pi G_N\beta}{3(2\epsilon-1)}}(\psi-\psi_0)\bigg]
\,.
\label{eq:avspsi}
\end{equation}
This means that --  just like $\phi$  -- in a power law expansion
$\psi$ corresponds to an $\epsilon$-dependent reparametrization of
the scale factor of the Universe. Note that $\epsilon = 1/2$ is a
singular point of the relation~(\ref{eq:avspsi}). This is not
unexpected, since from Eq.~(\ref{eq:herm:cosmology:3}) we know
that $\epsilon = 1/2$ corresponds to the case when $\dot\psi=0$
and hence also $W=0$, such that $a$ is given by~(\ref{avsphi}) and
does not dependent on $\psi$ (in fact in this case $\psi$ does not
even exist).

 More general cosmologies, with $\epsilon$ in Eq.~(\ref{eq:herm:cosmology:3b})
being a function of $z^0$ are possible, provided one chooses
$w(\psi)$ in Eq.~(\ref{Eq:W}) of a more general (non-exponential)
form. In these more general cosmologies no simple relation between
$\psi$ and $a$ exists such that Eq.~(\ref{eq:avspsi}) must be
suitably generalized.

\begin{figure}[t!]
  \centering
 \includegraphics[width=\columnwidth]{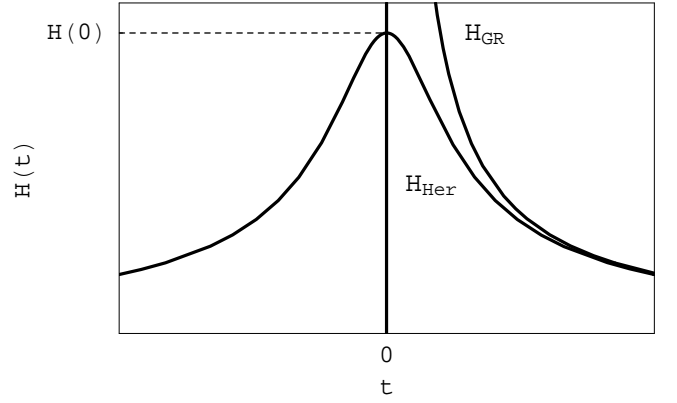}
  \caption{The observed expansion rate as a function of time.
  When moving backwards in time, the expansion rate of the Hermitian Hubble parameter ${\text H}_{\text{Her}}$ reaches a maximal value at $t=0$,
  whereas the Hubble parameter of general relativity ${\text H}_{\text{GR}}$ becomes infinite in finite time.}
  \label{fig:HubbleMax}
\end{figure}

\begin{figure}[t!]
  \centering
 \includegraphics[width=\columnwidth]{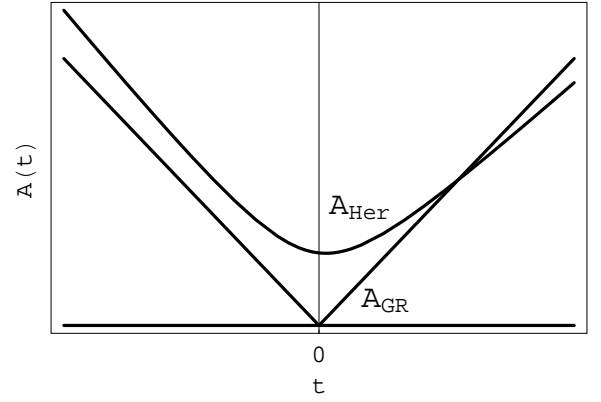}
  \caption{The observed scale factor as a function of time. The expansion rate reaches a minimal value at $t=0$.
   Nonzero $\text{Im}[\epsilon] $ breaks time reversal symmetry, $T$.}
  \label{fig:scalefactor}
\end{figure}

 Let us have a more careful look at the power law
solution~(\ref{eq:power law expansion}). Recall that in the
physical space an observer sees the expansion rate ${\cal H}$ that
can be obtained from the rotated Hermitian Einstein tensor
$G_{i\bar j}$~(\ref{eq:einstein t H cosmo unrotated}) as follows,
\begin{equation}
 {\cal H}^2 = -\frac{2}{45}\Re\bigg(\frac{G_{i\bar j}}{a\bar a}\bigg)
            = H\bar H
\,,
\nonumber
\end{equation}
from which we conclude (see Eq.~(\ref{eq:power law expansion})),
\begin{equation}
 {\cal H} = \frac{1}{|\epsilon|\sqrt{t^2+(G_NE/c^4)^2}}
\,.
\label{eq:H:physical}
\end{equation}
Note that at late times $t^2\gg G_N|E|/c^4$ the expansion rate
approaches that of general relativity,
\begin{equation}
 {\cal H} \rightarrow \frac{1}{|\epsilon|t}
\,,\qquad (t\rightarrow \infty)
\,,
\label{eq:H:physical:late}
\end{equation}
with $\epsilon_{GR}$ given by $|\epsilon|$ of Hermitian gravity.
In contrast to general relativity at early times $|t|\leq G_N
|E|/c^4$ the expansion rate does not diverge. Instead, it reaches
a maximal value at $t=0$ given by
\begin{equation}
  {\cal H} \rightarrow {\cal H}_{\rm max} =\frac{c^4}{|\epsilon|G_N|E|}
\,\qquad (t\rightarrow 0)
\,,
\label{eq:H:physical:early}
\end{equation}
which is nonsingular as long as $E\neq 0$ (below we  discuss the
physical relevance of the singularity at $t=0=E$). This behavior
of ${\cal H}$ corresponds to a bouncing cosmology. Indeed, since
the expansion rate is symmetric under time reversal, $t\rightarrow
-t$, for $t<0$ the Universe passes through a contracting phase,
followed by a mirror symmetric expanding phase for $t>0$. The time
dependence of ${\cal H}$ on time $t$ (on an $E={\rm const.}$
hypersurface) is shown in figure~(\ref{fig:HubbleMax}).
For completeness we now consider the scale factor of Hermitian
cosmology. The observed scale factor ${\cal A}$ corresponds to the
rotated metric tensor~(\ref{eq:rotated line
element:1}--\ref{eq:rotated line element:3}), $g_{\mu\nu}\equiv
{\cal A}^2(x,y)\eta_{\mu\nu} = \text{Re}[C_{\mu\bar\nu}]$. This
then implies,
\begin{eqnarray}\label{eq:A}
    {\cal A}
    =
    \sqrt{a\bar a}
    =
    {\cal A}_0 \left(\frac{\sqrt{t^2+(G_N E/c^4)^2}}{|\zeta_0|}
                           \right)^{(1/\epsilon)_R}
                           \times
  \\\nonumber
  \times \exp\bigg[\!-\!\Big(\frac{1}{\epsilon}\Big)_I
               \bigg({\rm Arctan}\Big(\frac{G_N E}{c^4t}\Big)
                 \!-\! \frac{\pi}{2}{\rm sign}(t)\bigg)
                     \Bigg]
\,,
\end{eqnarray}
where we chose $\zeta_0 = |\zeta_0|$ (the phase ${\rm
arg}(\zeta_0)$ can be absorbed in ${\cal A}_0$), $(1/\epsilon)_R =
\text{Re}[\epsilon]/|\epsilon|^2$, $(1/\epsilon)_I =
-\text{Im}[\epsilon]/|\epsilon|^2$, and we chose the Riemann sheet
of $a=a(z^0)$ such that ${\cal A}$ is continuous at $t=0$, as
required by the equations of motion for $a$. The observed scale
factor reaches a minimum at $t= 0$ and expands symmetrically under
time reversal, for $\text{Im}[\epsilon] = 0 $. Nonvanishing
$\text{Im}[\epsilon] $, however, violates charge-parity (CP)
symmetry of the $\psi$ field, as can be seen from the potential
(\ref{eq:V:power-law}). Since CPT is conserved, T must also be
violated. A manifestation of this T violation can be seen in
figure \ref{fig:scalefactor}, in which we can see that the
contracting phase is not a mirror image of the expanding phase.
The sign of $\text{Im}[\epsilon]$ determines the direction of the
tilt in the scale factor function. Note that ${\cal H}$ {\it
cannot} be obtained from ${\cal A}$ as $(\partial_t {\cal
A})/{\cal A}$. \footnote{
 Indeed, integrating ${\cal H}$ would result in the scale factor proportional to
\[
\Bigg(\frac{c^4t}{G_NE}+\sqrt{\Big(\frac{c^4t}{G_NE}\Big)^2+1}\;\Bigg)^{1/|\epsilon|}
\,,
\]
which differs from Eq.~(\ref{eq:A}).
}
This should not surprise us, given the fact that in Hermitian gravity
a derivative of a projected quantity onto a space-time hypersurface
is not in general equal to the projected derivative of the same
quantity. 
Mathematically, the difference arises because the projection
procedure must be made consistent with the Cauchy-Riemann
equations.

At late times $t\gg |G_N E/c^4|$ the solution~(\ref{eq:A})
approaches a power law expansion of general relativity,
\begin{eqnarray}\label{eq:A:2}
    {\cal A}
    \stackrel{t\rightarrow \infty}{\longrightarrow}
    \hat {\cal A}_0
    \left(\frac{t}{|\zeta_0|}
    \right)^{(1/\epsilon)_R}
\phantom{h}
    \hat{\cal A}_0
    =
    {\cal A}_0\exp\bigg[\frac{\pi}{2}
    \Big(\frac{1}{\epsilon}\Big)_I \bigg].
\end{eqnarray}
From this we see that at late times it is difficult to distinguish
the standard FLRW cosmology and Hermitian cosmology. Indeed, the
only difference is in the size of the Universe: when
$(1/\epsilon)_I = -\text{Im}[\epsilon]/|\epsilon|^2>0$
 ($(1/\epsilon)_I < 0$) the
Universe of Hermitian cosmology appears greater (smaller) than
the FLRW Universe with $\epsilon \leftrightarrow [(1/\epsilon)_R]^{-1}$.
Since the absolute value of the scale factor
cannot be observed (only ratios are observable),
this difference cannot be used to distinguish between
the standard and Hermitian cosmologies. If we had
any information about the size of the early Universe,
we could make the desired distinction.

One the other hand, at early times (when $t$ and $G_N E/c^4$ are
comparable), the two cosmologies differ quite dramatically.
Consider first the Universe which expands such that $E={\rm
const}$. In this case Eq.~(\ref{eq:A}) represents a bouncing
universe with a minimal size given by,
\begin{equation}
  {\cal A}_{\rm min} = {\cal A}(t=0) = {\cal A}_0 \left(\frac{G_N |E|}{c^4|\zeta_0|}
                           \right)^{(1/\epsilon)_R}
\,.
\label{eq:Amin}
\end{equation}
The Universe behaves regularly `everywhere' provided
the cosmological singularity at
$ E\rightarrow 0, t\rightarrow 0$
is never reached.

\subsection{Cosmological singularity}
\label{Cosmological singularity}

\noindent In order to find out how accessible the singular point
of Eqs.~(\ref{eq:H:physical}) and~(\ref{eq:A}) actually is, we
consider a freely falling observer, which falls `backwards in
time' towards the singularity
.
In order to study how velocities and energy change in an expanding
universe, we need to solve the corresponding geodesic equations.

 Let us begin with general relativity. We are working in a spatially flat FLRW space.
In conformal coordinates the Levi-Civit\`a connection is of the form,
\begin{eqnarray}\nonumber
    \Gamma^\mu_{\alpha\beta}
    =
    \frac{a'}{a}\Big(\delta^\mu_{\;\alpha}\delta_\beta^{\;0}
                    + \delta^\mu_{\;\alpha}\delta_\beta^{\;0}
                    + \delta^0_{\;\mu}\eta_{\alpha\beta}\Big).
\end{eqnarray}
This implies the following geodesic equation and line element (for
a massive observer),
\begin{eqnarray}\label{FLRW:geodesic eq}
  \frac{du_c^\mu}{d\tau}
    + \frac{a'}{a}\bigg(2u_c^0u_c^\mu
             - \frac{\delta^\mu_{\,0}}{a^2}\bigg)
   =
   0,
   \phantom{ha}
   \eta_{\alpha\beta}u_c^\alpha u_c^\beta
   =
   -\frac{1}{a^2},
\end{eqnarray}
where $\tau$ is the proper time observed by a freely falling observer
(in the frame in which all 3-velocities vanish): $(ds)^2=-(d\tau)^2$,
and $u^\mu_c =dx_c^\mu/d\tau$
is the 4-velocity in conformal coordinates $x_c^\mu = (\eta,x^i_c)$
(here we take $c=1$). The spatial equation~(\ref{FLRW:geodesic eq})
is easily solved,
\begin{equation}
  \frac{d (a^2 u_c^i)}{d\eta} = 0
\,,
\label{FLRW:geodesic eq:i}
\end{equation}
where we made use of the definition of conformal time, $u_c^0 d\tau = d\eta$.
This means that in an expanding universe, $u_c^i\propto 1/a^2$, such
that the physical momentum, $p^i_p = m a u_c^i$, scales as
$p^i_p\propto 1/a$, where $m$ is observer's mass. The time component of
Eq.~(\ref{FLRW:geodesic eq}) implies,
\begin{equation}
  \frac{d}{d\eta}\Big[a^2\Big(a^2(u_c^0)^2-1\Big)\Big] = 0
\,,
\label{FLRW:geodesic eq:0}
\end{equation}
which is consistent with the line element in Eq.~(\ref{FLRW:geodesic eq})
and with Eq.~(\ref{FLRW:geodesic eq:i}).
Eqs.~(\ref{FLRW:geodesic eq}) and~(\ref{FLRW:geodesic eq:i})
can be also used to determine the scaling of
the physical energy, $E_p \equiv p^0_p = m a u_c^0$:
\begin{equation}
 E_p^2 - \sum_i({p_p^i})^2 = m^2
\,,
\end{equation}
from where it follows, $E_p^2-m^2 \propto 1/a^2$
(this can be also concluded from Eq.~(\ref{FLRW:geodesic eq:0})).

\bigskip

\noindent Let us now consider Hermitian gravity. The relevant
connection coefficients are given in Eqs.~(\ref{eq:connection H
Cosmo unrotated}), such that the corresponding geodesic equation
(\ref{eq:geodesic equation arbitrary complex}) and the line
element~(\ref{eq:Hermitian familiar form line element}) are then,
\begin{eqnarray}\nonumber
  \frac{du_c^\mu}{d\tau}
    + \frac{a'}{a}u_c^0u_c^\mu
    + \frac{\bar{a}'}{\bar{a}}\Big(u_c^{\bar 0}u_c^{\mu}
    &+&
    \eta^{\mu\bar 0}\frac{1}{a\bar a}\,\Big)
   = 0
   \\\label{FLRW:geodesic eq:her}
   \eta_{\alpha\bar\beta}u_c^\alpha u_c^{\bar \beta}
   &=&
   -\frac{1}{a\bar a}
\end{eqnarray}
where again $\tau$ is a real affine parameter defined as the
proper `time' of a freely falling observer (in the frame in which
all 3-velocities vanish and $E=0$): $(ds)^2=-2(d\tau)^2$, and
$u^\mu_c =dz_c^\mu/d\tau$ is the complex proper 4-velocity in
conformal coordinates $z_c^\mu = (z^0_c,z^i_c)$. Notice that from
the definition $dz_c^0/d\tau = u_c^0$, it follows that the second
term in Eq.~(\ref{FLRW:geodesic eq:her}) can be absorbed by a
simple rescaling of $u_c^\mu$, such that it simplifies to
\begin{eqnarray}\nonumber
  \frac{d(au_c^\mu)}{d\tau}
    + \bar H \Big((\bar au_c^{\bar 0})(au_c^{\mu})
    +
    \eta^{\mu\bar 0}\,\Big)
    &=& 0
\\\label{FLRW:geodesic eq:her:2}
    \eta_{\alpha\bar\beta}(au_c^\alpha)(\bar a u_c^{\bar \beta})
    &=& -1 ,
\end{eqnarray}
where $\bar H = \bar a'/{\bar a}^2 \equiv \dot {\bar a}/\bar a$,
 $dz^0=adz_c^0$ and $\dot {\bar a} = (1/a)da/dz_c^0$. When split into components Eq.~(\ref{FLRW:geodesic eq:her:2}) yields,
\begin{subequations}
\begin{eqnarray}
  \frac{du^0}{d\tau}
    &+& \bar H
   \big(u^{\bar 0}u^{0} - 1\big)
   = 0
\label{FLRW:geodesic eq:her:3a}
\\
  \frac{du^i}{d\tau}
    &+& \bar H u^{\bar 0}u^{i}
   = 0
   \,,\qquad u^0u^{\bar 0}-u^iu^{\bar i}
               = 1
\,,
\label{FLRW:geodesic eq:her:3b}
\end{eqnarray}
\end{subequations}
where we defined the complex `physical'
4-velocities $u^\mu=au_c^\mu$ and $u^{\bar \mu}=\bar au_c^{\bar \mu}$.
The corresponding complex conjugate equations must also hold.
The temporal equation~(\ref{FLRW:geodesic eq:her:3a})
and its complex conjugate can be combined to give,
\begin{equation}
 \frac{d}{d\tau}\ln\big(u^0u^{\bar 0}-1\big) = - \big(H u^0+\bar Hu^{\bar 0}\big)
       = - \frac{d\ln(a\bar a)}{d\tau}
\,.
\label{FLRW:geodesic eq:her:4}
\end{equation}
The last equality follows from
$H u^0 = d\ln(a)/d\tau$ and $\bar H u^{\bar 0} = d\ln(\bar a)/d\tau$.
This can be straightforwardly integrated from $\tau_0$ to $\tau$
resulting in the scaling,
\begin{equation}
  \frac{u^0u^{\bar 0}-1}{(u^0u^{\bar 0})_0-1} =
                             \frac{(a\bar a)_0}{a\bar a}
                          = \frac{u^iu^{\bar i}}{(u^iu^{\bar i})_0}
\,,
\label{FLRW:geodesic eq:her:sol:1}
\end{equation}
where the last equality follows from the constraint in
Eq.~(\ref{FLRW:geodesic eq:her:3a}), or equivalently from
Eq.~(\ref{FLRW:geodesic eq:her:3b}).
In Eq.~(\ref{FLRW:geodesic eq:her:sol:1})
$u^\mu=u^\mu(\tau)$, $a=a(\tau)$ and we have defined
$(u^0u^{\bar 0})_0 = u^0(\tau_0)u^{\bar 0}(\tau_0)$,
$(u^iu^{\bar i})_0=u^i(\tau_0)u^{\bar i}(\tau_0)$
and $(a\bar a)_0=a(\tau_0)\bar a(\tau_0)$.
Analogously to general relativity the spatial components
of particles' physical (complex) velocities scale as,
\begin{equation}
 u^iu^{\bar i} = \Big(\frac{dx^i}{d\tau}\Big)^2
                    + G_N^2 \Big(\frac{dp^{\check i}}{d\tau}\Big)^2
            \propto \frac{1}{a\bar a} = \frac{1}{{\cal A}^2}
\,.
\end{equation}

Next we recall that,
\begin{equation}
  u^0 = \frac{dz^0}{d\tau}
\,,\qquad
  u^{\bar 0} = \frac{dz^{\bar 0}}{d\tau}
\nonumber
\end{equation}
and we define a radial and angular (time-like) coordinates,
\begin{equation}
  z^0 \equiv r{\rm e}^{i\theta} = \frac{x^0+iy^0}{\sqrt{2}}
\,.
\nonumber
\end{equation}
Now, by making use of the definition
$u^0=(d/d\tau)(r{\rm e}^{i\theta})$ one immediately arrives at
the identity,
\begin{equation}
 {\cal E}\equiv \frac12{\dot r}^2 + V(r,\theta) = 0
\,,\qquad
   V = \frac12\frac{L^2}{r^2} - \frac12u^{0}u^{\bar 0}
\,,
\label{FLRW:geodesic eq:her:6}
\end{equation}
where we defined an `angular momentum'
\begin{equation}
L=r^2\dot\theta
\,.
\label{eq:L:def}
\end{equation}
This angular momentum (or more precisely the angular velocity
$\omega_\theta=\dot\theta$) characterizes the rate of mixing
between the time and energy coordinates in Hermitian gravity.
Equation~(\ref{FLRW:geodesic eq:her:6}) represents the conserved
`energy density' of Hermitian cosmology. Indeed, since ${\cal
E}=0$, the energy density~(\ref{FLRW:geodesic eq:her:6}) is
trivially conserved, $\dot {\cal E}=0$. The angular
momentum~(\ref{eq:L:def}) is, however, not generally conserved,
implying that the time and energy generally mix. This can be seen
from the imaginary part of Eq.~(\ref{FLRW:geodesic eq:her:3a})
which -- when divided by $\bar\epsilon \bar H$ and using the
Hubble parameter~(\ref{eq:power law expansion}) of power law
expansion -- yields,
\begin{equation}
  \dot L \equiv \frac{d}{d\tau}(r^2\dot \theta)
                 = - \frac{\epsilon_I}{|\epsilon|^2}(u^0u^{\bar 0}-1)
\,.
\label{FLRW:geodesic eq:her:5}
\end{equation}

Requiring that the derivative of energy
integral~(\ref{FLRW:geodesic eq:her:6})
vanishes, one obtains the equation of motion for $r$,
which corresponds to the real part of Eq.~(\ref{FLRW:geodesic eq:her:3a})
(divided again by $\bar\epsilon \bar H$).
That means that Eq.~(\ref{FLRW:geodesic eq:her:6}) is an integral of motion and
remarkably the dynamics of particles in Hermitian cosmology reduces
to a study of motion in a (simple) potential given
in Eq.~(\ref{FLRW:geodesic eq:her:6}).

\begin{figure}[ht]
  \centering
    \subfigure[ The potential as a function of the modulus $r$ for $\epsilon >1$. ]{
        \includegraphics[width=\columnwidth]{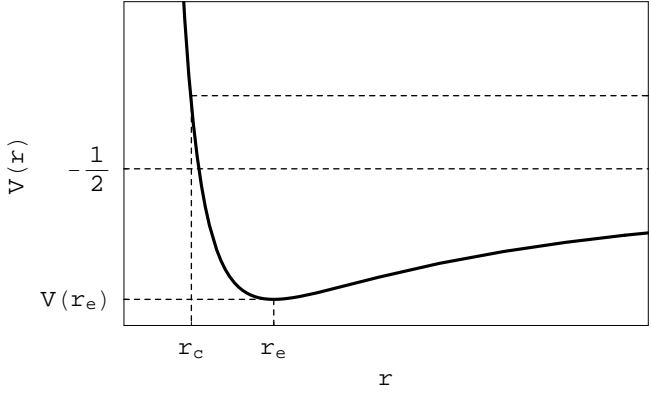}
        \label{fig:HermitianCosmology1}
    }
    \subfigure[  The potential as a function of the modulus $r$ for
        $0<\epsilon<1$. Whenever $V(r_c) > 0$ there is a bounce.]{
        \includegraphics[width=\columnwidth]{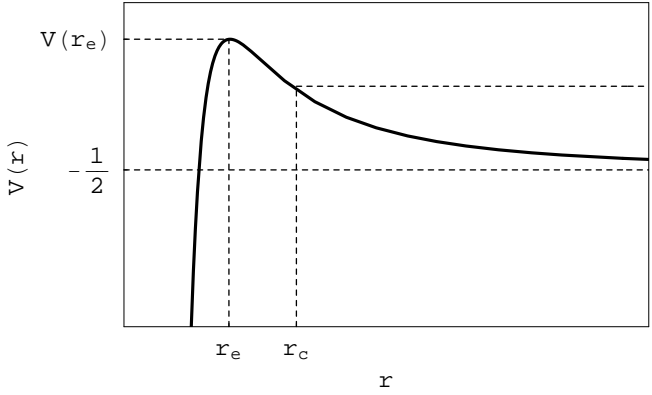}
        \label{fig:HermitianCosmology2}
    }
    \subfigure[  The potential as a function of the modulus $r$ for
        $\epsilon = 1$. For $L_0^2>U_0$ there is a bounce.]{
        \includegraphics[width=\columnwidth]{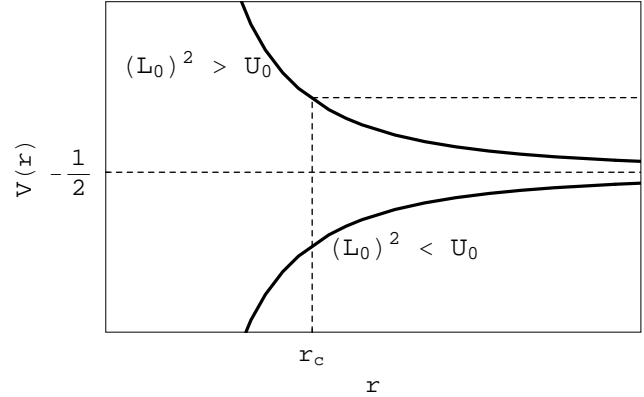}
        \label{fig:HermitianCosmology3}
    }
    \label{fig:HermitianCosmology}
    \caption{Potential (\ref{eq:potential:epsReal}) as a function of the radius for different values of
    $\epsilon$ exhibits quite generically a bounce cosmology.}
\end{figure}

 In order to illustrate how to completely solve the geodesic equations of Hermitian
cosmology, we now restrict ourselves to the simple case when
$\epsilon_I=0$ (recall that in standard FLRW cosmology $\epsilon$
is by definition a {\it real} parameter). In this case
Eq.~(\ref{FLRW:geodesic eq:her:5}) implies that the angular
momentum $L=L_0$ is conserved, $\dot L_0=0$ and the
potential~(\ref{FLRW:geodesic eq:her:6}) acquires the simple form,
\begin{equation}
  V = \frac12 \frac{L_0^2}{r^2} - \frac{U_0}{2r^{{2}/{\epsilon}}} - \frac12
\,,\qquad U_0 = [(u^{0}u^{\bar 0})_0-1]
       (a\bar a)_0 |\zeta_0|^{{2}/{\epsilon}}
\,,
\label{eq:potential:epsReal}
\end{equation}
where $U_0\geq 0$ parameterizes the time-like velocity at a time
$\tau_0$. Provided $\epsilon\neq 1$ this potential has an extremum
$V_e$
  at the radius $r_e$ given by,
\begin{eqnarray}\nonumber
   V_e &=& - \frac{U_0}{2}\frac{\epsilon-1}{\epsilon}
           \left(\frac{U_0}{\epsilon L_0^2}\right)^\frac{1}{\epsilon-1}
               -\frac12
\\\label{re:Ve}
  r_e &=& \left(\frac{\epsilon
  L_0^2}{U_0}\right)^\frac{\epsilon}{2(\epsilon-1)}
 \phantom{ha}(L_0\neq 0, U_0\neq 0).
\end{eqnarray}
When $\epsilon>1$ (decelerated expansion) the extremum is a
minimum, as can be seen from figure \ref{fig:HermitianCosmology1}.
Whenever $0<\epsilon<1$ (accelerated expansion) the extremum is a
maximum, as is depicted in figure \ref{fig:HermitianCosmology2}.
Choosing $\epsilon = 2$ corresponds to the radiation era and the
value $\epsilon = 1/2$ is close to the $\epsilon$ parameter of
today's Universe. The critical value of $\epsilon= 1$ yields a
curvature dominated universe, as is shown in figure
\ref{fig:HermitianCosmology3}. In this case V does not have an
extremum (formally, an extremum $V\rightarrow -1/2$ is reached for
$r\rightarrow \infty$). As $r\rightarrow 0$ the potential
approaches $+\infty$ ($-\infty$) when $L_0^2>U_0$ ($L_0^2<U_0$).

  Since the energy integral~(\ref{FLRW:geodesic eq:her:6}) is conserved,
particle dynamics in the potential~(\ref{eq:potential:epsReal})
with $\epsilon>1$ (or when $\epsilon=1$ and $L_0^2>U_0$) is such
that they reach the minimal distance $r_c$ (a turning point) given
by the point where $\dot r(r_c)=0$. In other words, the Universe
of Hermitian gravity generically exhibits a bounce whenever
$\epsilon>1$ and $L_0\neq 0$. The critical (minimal) radius is
given by $V(r_c)=0$, as can be seen from (\ref{FLRW:geodesic
eq:her:6}). In the radiation era ($\epsilon=2$), we can solve
analytically for the critical radius
\begin{eqnarray}\nonumber
    r_c
    \equiv
    \sqrt{t_c^2 + (G_NE)^2} = \Delta -\frac12 U_0 \,,
\\\label{eq:rc}
    \Delta
    =
    \sqrt{(U_0/2)^2+L_0^2} \qquad ({\rm radiation}\;\;{\rm
    era}) \,.
\end{eqnarray}

The integral from of Eq.~(\ref{FLRW:geodesic eq:her:6}) is,
\begin{equation}
 \int \frac{dr}{\sqrt{-2V}}=\pm \tau
\,,
\label{eq:general integral rvstau}
\end{equation}
which cannot be performed analytically for a general $\epsilon$.
In radiation era ($\epsilon = 2$) integrating~(\ref{eq:general integral rvstau})
gives,
\begin{eqnarray}\label{rvstau}
    &\pm& \tau
    =
    \sqrt{r^2 + U_0r-L_0^2}
    \\\nonumber
    &-&\frac{U_0}{2}\ln \Bigg(\frac{r+({U_0}/{2})+\sqrt{r^2 + U_0r-L_0^2}}
                     {\Delta}\Bigg)
\,,
\end{eqnarray}
where we chose the proper time $\tau$ such that $r(\tau=0)=r_c$.
This can be inverted close to the bounce,
\begin{equation}
  r \simeq r_c + \frac{\Delta}{2}\Big(\frac{\tau}{r_c}\Big)^2
\,,\qquad (r-r_c\ll \Delta,\tau\gg r_c)
\nonumber
\end{equation}
At late times one gets the expected linear behavior plus a
logarithmic correction which characterizes Hermitian gravity,
\begin{equation}
  r \simeq \tau + \frac{U_0}{2}\bigg[\ln\Big(\frac{\tau}{\Delta}\Big)-1\bigg]
\,,\qquad (r-r_c\gg \Delta,\tau\gg r_c)
\,.
\nonumber
\end{equation}
In the limiting case when $\epsilon = 1$
(curvature domination) Eqs.~(\ref{FLRW:geodesic eq:her:6})
and~(\ref{eq:potential:epsReal}) can be integrated to give,
\begin{equation}
  \sqrt{r^2 + U_0 -L_0^2} = \pm \tau
\label{eq:eps=1}
\end{equation}
such that when $U_0<L_0^2$ there is a bounce with the minimal
Hubble length given by,
\begin{equation}
  r_c = \sqrt{L_0^2 -U_0} \,,\qquad (\epsilon=1,\,L_0^2>U_0)
\,.
\label{eq:rc:eps=1}
\end{equation}

The existence of a minimal Hubble length $r_c$ as given by
Eqs.~(\ref{eq:rc}) and~(\ref{eq:rc:eps=1}) means that even when
time $t_c$ is set to zero (`Big Bang'), the Universe reaches its
maximal -- but finite -- expansion rate~(\ref{eq:H:physical})
\begin{equation}
 {\cal H}_{\rm max} = \frac{1}{\epsilon r_c}
                     = \frac{1}{\epsilon\sqrt{t_c^2+(G_NE_c/c^4)^2}}
\,,
\label{eq:H:physical:max}
\end{equation}
with $r_c$ given in Eq.~(\ref{eq:rc}).
(Even if $L_0$ were set to zero initially, a small nonvanishing $\epsilon_I$
would violate angular momentum conservation~(\ref{FLRW:geodesic eq:her:5}),
such that we expect that $L\neq 0$ generically close to the bounce.
Moreover, the choice $L_0=0$ represents a set of measure zero in the space of
all initial conditions $\{u^\mu(\tau_0)|
      \eta_{\mu\bar\nu}u^\mu(\tau_0)u^{\bar\nu}(\tau_0)=-1\}$, and in this sense
the condition $L_0=0$ is `almost never' realized.)
Equation~(\ref{eq:H:physical:max}) constitutes the main result of
our analysis of Hermitian cosmology, according to which Hermitian
cosmology is nonsingular at the classical level.

Note that Hermitian cosmology predicts $r_c$, but at what time,
$t_c$, and energy, $E_c$, $r_c$ is reached depends on the initial
conditions embodied by $U_0$ and $L_0$. In other words: $\theta_c$
is not predicted since the corresponding angular velocity is
associated with a conserved quantity $L_0$. To see this let us
consider the evolution of the mixing angle $\theta$,
\begin{equation}
  \theta = \theta_0 + \int_{r_0}^r\frac{Ldr}{r^2\sqrt{-2V}}
\,.
\label{eq:theta:general}
\end{equation}
This can be integrated for example in radiation era
($\epsilon=2$). This is the case, since $\epsilon_I=0$, $L=L_0$ is
conserved, and the integral~(\ref{eq:theta:general}) evaluates to
\begin{equation}
  \theta = \theta_0
  -{\rm Arcsin}\bigg(\frac{\frac{L_0^2}{r}-\frac{U_0}{2}}{\Delta}\bigg)
\,,\qquad \Delta^2 = L_0^2 + \Big(\frac{U_0}{2}\Big)^2
\,,
\label{eq:theta:radiation}
\end{equation}
where we absorbed the value of the integral at $r_0$ into
$\theta_0$. Because of the undetermined $\theta_0$,
$\theta_c\equiv\theta(r_c)$ is indeed not predicted. Yet demanding
$\theta\rightarrow 0$ when $r\rightarrow\infty$ gives $\theta_0 =
-{\rm Arcsin}(U_0/(2\Delta))$. At the minimal radius $r=r_c$,
Eq.~(\ref{eq:theta:radiation}) implies $\theta_c=\theta_0 -\pi/2$,
such that $\theta_c$ can be anywhere between $-\pi$ and $\pi/2$,
depending on $L_0$ and $U_0$. For example, in the limit when
$L_0/U_0\rightarrow 0$, $\Delta \theta \rightarrow -\pi$, while in
the opposite limit when $L_0/U_0\rightarrow \infty$, $\Delta
\theta \rightarrow -\pi/2$. Note that in the latter case the
Universe's expansion rate at the minimal radius $r_c$ is
completely determined by $E_c$.

 To complete the analysis of the geodesic equation,
one needs to integrate Eq.~(\ref{FLRW:geodesic eq:her:3b}).
By observing that $\bar H u^{\bar 0}=d\ln(\bar a)/\tau$, one integral
can be trivially performed, resulting in
\begin{equation}
  \frac{dz^i}{d\tau} = u^i(\tau_0)\frac{\bar a_0}{\bar a(\tau)}
\,.
\nonumber
\end{equation}
This can be integrated to get $z^i=z^i(\tau)$
in special cases by making use of the dependence of
the scale factor $a=a(r,\theta)$ in Eq.~(\ref{eq:A}), based on which
analysis of the causal structure of Hermitian cosmology
can be performed. We postpone this analysis for future work.

 When $\epsilon<1$ and when $V_e<0$ in Eq.~(\ref{re:Ve})
(or when $\epsilon=1$ and $L_0^2<U_0$)
the Universe collapses towards the Big Bang singularity
$r\rightarrow 0$ in a finite time. This will be the case
only when the weak energy condition is violated,
that is when $\rho+3p<0$, where  $\rho$ and $p$
denote the energy density and pressure of the cosmological fluid, respectively.
(These statements are based on the relation, $\epsilon = (3/2)(1+w)$, where $w=p/\rho$,
which holds in standard FLRW cosmology.)
Notice that even when $\epsilon<1$, the Universe may exhibit
a bounce, provided $V_e>0$, or equivalently if the angular momentum
is large enough,
$L_0^2>U_0^{2-\epsilon}\epsilon^{-\epsilon}(1-\epsilon)^{-(1+\epsilon)}$.
In this case there is a finite barrier for a Universe to tunnel to
smaller radii where $V(r)<0$; if that happens, the Universe hits eventually
the Big Bang singularity ${\cal H}\rightarrow \infty$.
This means that inflation and bounce cosmology are not
mutually incompatible.

 We have thus shown that Hermitian gravity solves the problem of Big
Bang singularity of Einstein's theory in a natural way.

\section{The cosmological constant problem}
\label{The cosmological constant problem}

Let us first recall Eqs.~(\ref{eq:Hermitian Einstein's
equations:1}--\ref{eq:Hermitian Einstein's equations:2}), which we
now write as,
\begin{subequations}
\begin{eqnarray}
  G_{\mu\nu} + C_{\mu\nu} \Lambda   &=& 8\pi G_N T_{\mu\nu}
\label{eq:HEeqs:1a}
\\
  G_{\mu\bar\nu} + C_{\mu\bar\nu}\Lambda &=& 8\pi G_N T_{\mu\bar\nu}
\,. \label{eq:HEes:2a}
\end{eqnarray}
\end{subequations}
Now imposing the reciprocity symmetry on shell implies
\begin{equation}
  C_{\mu\nu} = 0
\,,
\end{equation}
which means that the geometric cosmological term cannot contribute
to the holomorphic equation~(\ref{eq:HEeqs:1a}). Furthermore, as
we have seen in section~\ref{Hermitian Cosmology}, the reciprocity
symmetry reduces the Hermitian sector~(\ref{eq:HEes:2a}) to the
constraints~(\ref{eq:herm:cosmology:1}--\ref{eq:herm:cosmology:2}).

 A simple proof that these constraints
cannot be met unless $\Lambda$ is fully compensated by a constant
term in the scalar potential follows from the observation that the
form of the scalar potential $V=V(\phi,\psi)$ is uniquely given by
Eq.~(\ref{eq:V:1}), with $\Omega = \pm\sqrt{4\pi G_N\alpha/3}$,
$\lambda = \Lambda/(8\pi G_N)$ and $W=W(\psi,\bar\psi)$. Note that
the term  $-\Lambda/(8\pi G_N)$ in the potential~(\ref{eq:V:1})
cancels exactly the geometric cosmological constant $\Lambda$ in
Eq.~(\ref{eq:HEes:2a}). This proof applies only to Hermitian
cosmology governed by two scalar fields $\phi$ and $\psi$ as
described by Eqs.~(\ref{eq:complex action:full}--\ref{eq:scalar
action:fields}).

 Since this is an important point, we shall now construct an alternative proof,
which shows that the assumption that the late times Universe
approaches a de Sitter phase with a constant expansion rate
governed by some $\Lambda_{\rm eff}>0$ leads to contradiction,
resolved by requiring $\Lambda_{\rm eff}\rightarrow 0$.

Before we proceed, let us recall the standard FLRW cosmology
filled with a matter with an equation of state,
 $w_M=p/\rho>-1$ ($\epsilon_M = (3/2)(1+w_M)$)
 and a cosmological term $\Lambda$. The (classical) Hubble parameter
is of the form,
 \begin{subequations}
 \begin{eqnarray}
   H_{GR} = \sqrt{\frac{\Lambda}{3}}
               \coth\bigg(\epsilon_M\sqrt{\frac{\Lambda}{3}}\,t\bigg)
 \\ \label{Hcl}
   \dot H_{GR} = \frac{\Lambda}{3}\frac{\epsilon_M}
                 {\cosh^2\Big(\epsilon_M\sqrt{\frac{\Lambda}{3}}\,t\Big)}
\,,
\end{eqnarray}
\end{subequations}
such that at late times $t\gg (1+w_M)^{-1}\sqrt{\Lambda/3}$, the
expansion rate $H_{GR}$ approaches the de Sitter attractor,
$H_{GR}\rightarrow H_{dS}=\sqrt{\Lambda/3}$, and $\dot
H_{GR}\rightarrow 0$ exponentially fast. This means that a
universe filled with any matter with an equation of state with
$w_M>-1$ will eventually approach the late time de Sitter
attractor. This is the case, simply because the energy density in
any matter fluid, with $w_M>-1$, dilutes as $\rho_M\propto
1/a^{3(1+w_M)}\propto 1/t^2$ as the Universe expands (provided
$w_M$ is constant), such that at sufficiently late times the
cosmological constant necessarily dominates.

 To construct an alternative proof,
let us assume that at late times the Universe approaches a
solution with a non-zero effective cosmological constant
$\Lambda_{\rm eff}$, which yields a constant expansion rate
$H\rightarrow H_{dS}=\sqrt{\Lambda_{\rm eff}/3}$ and $\dot
H\rightarrow 0$. $\Lambda_{\rm eff}$ is not necessarily the
original geometric cosmological constant, yet it must be strictly
positive and $\partial_{z^0} \Lambda_{\rm eff}\rightarrow 0$ as
$|z^0|\rightarrow \infty$. Firstly, from
Eq.~(\ref{eq:herm:cosmology:1}) we see that as $|z^0|\rightarrow
\infty$,
\begin{eqnarray}\nonumber
    H\bar H
    =
    \frac{1}{9}\big(8\pi G_N V + \Lambda\big)
       \rightarrow \frac{\Lambda_{\rm eff}}{9} = {\rm const.}
\\\label{eq:HbarH}
    (\Lambda_{\rm eff} = 8\pi G_N V_{\rm eff}) \,,
\end{eqnarray}
or equivalently,
\begin{eqnarray}\nonumber
    (\partial_{z^{0}}H)\bar H
      \rightarrow \frac{8\pi G_N }{9} \partial_{z^0}V_{\rm eff}&\rightarrow& 0
\\ \label{eq:HbarH:dot}
    H\partial_{z^{\bar 0}}{\bar H}
     \rightarrow \frac{8\pi G_N}{9}\partial_{z^{\bar 0}}V_{\rm eff}
           \rightarrow 0
    \phantom{ha} (|z^0|&\rightarrow& \infty) \,.
\end{eqnarray}
 Next, we multiply Eq.~(\ref{eq:herm:cosmology:4}) by $\dot{\bar \phi}$
and make use of Eq.~(\ref{eq:herm:cosmology:1}) to arrive at,
\begin{equation}
  \dot\phi \dot{\bar \phi} \rightarrow
     - \frac{1}{3\alpha}\frac{1}{\bar H}
      \frac{\partial V_{\rm eff}}{\partial z^{\bar 0}}
\,, \label{eq:dotphi-dotphi}
\end{equation}
where we made use of $\dot{\bar \phi}\partial_{\bar \phi} V
 \rightarrow \dot{\bar \phi}\partial_{\bar \phi} V_{\rm eff}
 \equiv \partial V_{\rm eff}/\partial z^{\bar 0} $, with
$V_{\rm eff} = \Lambda_{\rm eff}/(8\pi G_N)$. Now combining
Eqs.~(\ref{eq:dotphi-dotphi}) with Eq.~(\ref{eq:herm:cosmology:2})
yields,
\begin{equation}
 \frac{\partial \ln[V_{\rm eff}]}{\partial z^{\bar 0}} = -\frac12 \bar H
\,. \label{eq:constraint:z0 derivative}
\end{equation}
The analogous complex conjugate equation also holds. But from
Eq.~(\ref{eq:HbarH:dot}) we know that at late times $V_{\rm eff}$
must approach a constant, and thus
\begin{equation}
 \frac{\partial \ln[V_{\rm eff}]}{\partial z^{\bar 0}}
 \rightarrow 0
\,,\qquad
 \frac{\partial \ln[V_{\rm eff}]}{\partial z^{0}}
 \rightarrow 0
\,\qquad (|z^0|\rightarrow \infty) \,,
\end{equation}
implying finally that at late times $\bar H\rightarrow 0$, which
together with Eq.~(\ref{eq:HbarH}) gives,
\begin{equation}
  H\bar H \rightarrow \frac{\Lambda_{\rm eff}}{9} \rightarrow 0
\,\qquad (|z^0|\rightarrow \infty) \,. \label{eq:HbarH:2}
\end{equation}
This completes the proof that there is no late time de Sitter
attractor driven by a nonvanishing effective cosmological term
$\Lambda_{\rm eff}>0$ in Hermitian gravity (with the two scalar
field action~(\ref{eq:scalar action}) used in this article).

To summarize, we have shown that the consistency of Hermitian
gravity constraints requires $\Lambda_{\rm eff} = \Lambda + 8\pi
G_N V_0 \rightarrow 0$, where $V_0$ represents the time (and
energy) independent part of the scalar potential $V$. In other
words, any cosmological term of Hermitian gravity must be fully
and precisely compensated by the corresponding scalar potential.

 The question is whether this holds more generally when other types of
matter fields (fermions and gauge fields) are included. And
moreover, what happens when quantum corrections are included. We
postpone the discussion of these (important) questions for future
work.

  Nevertheless, note that an appropriate choice of the potential
for the second scalar field $\psi$ can lead to arbitrary (power
law) expansion rate, which also includes a near exponential
expansion with $\epsilon\simeq 0$. Even though this type of
conformal scalar $\psi$ matter behaves similar to a cosmological
term, it is not completely identical. In fact, the choice
$\epsilon = 0$ in Eq.~(\ref{eq:V:power-law}) is very particular
(it entails {\it fine tuning}), and thus does not comprise a
cosmological constant problem. Let us now consider the limit
$\epsilon\rightarrow 0$. The
potential~(\ref{eq:w:power-law}--\ref{eq:V:power-law})
\begin{equation}
w(\psi,\bar\psi) \rightarrow w_0
              \exp\Big[ i \omega\sqrt{(\beta/\alpha)}\, (\psi-\bar \psi)\Big]
\end{equation}
is oscillatory (here we used $\sqrt{-1}=i$). This potential
becomes exponential if $\beta/\alpha<0$.

  Note also that $\epsilon=1/2$ ($w=-2/3$) has special relevance.
This power-law accelerated expansion is realized in the absence of
the second field $\psi$.

Let us now rewrite Eq.~(\ref{eq:power law expansion}) as,
\begin{equation}
 a = a_0\left(1+h\epsilon z^0\right)^{1/\epsilon}
\label{eq:a:shifted}
\end{equation}
where we shifted time $z^0\rightarrow z^0 + \zeta_0$ and we
defined, $h=1/(\epsilon \zeta_0)$. Now upon taking the limit
$\epsilon\rightarrow 0$, Eq.~(\ref{eq:a:shifted}) reduces to,
\begin{equation}
   a = a_0\exp\left(h z^0\right)
\,\qquad
  (h \in \mathbb{C})
   \,,
\label{eq:a:shifted:2}
\end{equation}
representing an (exponentially expanding) complex de Sitter
universe of Hermitian gravity with the complex Hubble parameter,
\begin{equation}
  H = h
\,. \label{eq:H:dS}
\end{equation}
This holomorphic de Sitter space must be distinguished from the de
Sitter space induced by a (real) cosmological term in the
Hermitian sector of the theory.

 In summary, we found that, as a consequence of the reciprocity symmetry,
Hermitean gravity does not admit a cosmological term at the
classical level neither in the holomorphic sector nor in the
Hermitean sector of the theory. Yet it does admit a holomorphic de
Sitter space realized by a holomorphic scalar field (with a
holomorphic kinetic term and with a suitably fine tuned
exponential potential).

 We have thus formulated a generalized theory of gravitation
which (at the classical level) allows Minkowski space, but does
not admit de Sitter space realized by a positive cosmological
constant.

\section{Discussion}
\label{Discussion}

\noindent We have formulated a generalized theory of gravity on
Hermitian manifolds. Given the extensive literature on complex
manifolds, we summarize (and emphasize) the novel aspects of our
work and compare it to existing literature:

 \begin{itemize}

\item[{\tt 1.}] Our Hermitian theory of gravity
 lives on a Hermitian manifold of real dimension eight. There
 are four space-time ($x^\mu$)
 and four momentum-energy ($p^\mu$) coordinates.
 The fundamental dynamical quantity of the theory
 is a holomorphic tetrad, which is a function of
 $z^\mu = x^\mu + i(G_N/c^3)p^\mu$.
 The tetrad transforms by means of
 holomorphic coordinate
transformations~(\ref{tetrad:coord.transformations}), which in
general
 mix space-time and momentum-energy. This extends and generalizes
both the principle of covariance and equivalence of general
relativity. We identify the reciprocity symmetry with the
operation of the almost complex structure operator, which
transforms $\partial/\partial x^\mu$ into $\partial/\partial
y^\mu$ and $\partial/\partial y^\mu$ into $-\partial/\partial
x^\mu$. The reciprocity symmetry of a Hermitian manifold demands
that the world on `very large scales' ($\ell \gg l_{\rm Pl}$)
(general relativistic limit) mirrors the world on `very small
scales' ($\ell \ll l_{\rm Pl}$) (microscopic super-Planckian
world), but with the role of space-time and momentum-energy
exchanged~\cite{Born1938Reciprocity}. This symmetry leaves the
commutation relation~(\ref{eq:commutation relations}) invariant.
The reciprocity symmetry implies holomorphy of the tetrad fields.
Holomorphy reduces the degrees of freedom of an eight dimensional
theory to the degrees of freedom of an effectively four
dimensional world, as required by all known observations.

 \item[{\tt 2.}] The eight dimensional formulation of the theory is
symmetric, and yet the (four dimensional) metric contains an
antisymmetric tensor, which corresponds to the imaginary part of
the metric tensor $C_{\mu\bar\nu}$~(\ref{eq:Einstein's
decomposition of the Hermitian line element}), which gives rise to
dynamical torsion (this still awaits a rigorous proof). Our theory
differs from other dynamical theories of torsion (see for
example~\cite{Moffat:1978tr}) in the leading order dynamics of
torsion. It results from a theory that is projected on the
space-time submanifold such that different orders in $p^\mu$ mix
as a consequence of the Cauchy-Riemann equations (see remark {\tt
6} below). In this work we do not address the dynamics of the
antisymmetric part of the metric, which may be of importance for
example for the dark matter of the
Universe~\cite{Prokopec:2006kr}, for spinning black holes and for
the Lens-Thirring effect. Yet the fact that our Hermitian gravity
theory corresponds to a `standard' gravity theory of a symmetric
metric field on an eight dimensional (Hermitian) manifold, is a
strong indication that the theory of torsion within our Hermitian
gravity does not suffer from the stability
problems~\cite{Damour:1991ru,Janssen:2006jx} of -- for example --
the NGT of Ref.~\cite{Moffat:1978tr}.

 \item[{\tt 3.}] We define parallel transport by means of
a metric compatible covariant derivative $\nabla_\mu$.
 Contrary to most (mathematical) literature on Hermitian
 manifolds~\cite{Nakahara:2003nw},
our covariant derivative is metric compatible,
 but {\it not} tetrad compatible (\ref{eq:vielbein compat}) in the sense discussed. We consider our definition
 of the covariant derivative as more natural and better
 physically motivated, as it stems from
 the action principle for test particles, $S=-m\int ds$.
Our covariant derivative implies `nonstandard' Hermitian
connection coefficients~(\ref{eq:Hermitian connection coefficients
z basis 4 d}).

  \item[{\tt 4.}] The causal structure of the theory is changed
such that in the flat space limit, the space-time-momentum-energy
line element is invariant under the
$U(1,3)$ group (the Hermitian line element is also invariant under
complex translations and hence invariant under the Hermitian  generalization of the Poincar\'e group). 
The momentum-energy coordinates can be interpreted as coordinates
describing non-inertial frames. The $U(1,3)$ reduces to its
subgroup, the Lorentz group $SO(1,3)$, whenever observers move
inertially with respect to each other; the momentum energy part of
the Hermitian flat space line element vanishes.
When observers move non-inertially with respect to each other, the
principle of covariance of general relativity is broken, but at
the same time replaced by an extended principle of covariance
(namely, the flat space metric is invariant under the U(1,3)
group). Yet this breaking becomes significant only in strong
gravitational fields and for large momenta and energies of
observers/particles, and hence does not necessarily contradict
observations.

The causal structure of the flat space limit is changed in such a
way that there is a minimal time for events to be in causal
contact and a maximal radius $r_{\rm max}$ for a non-local
instantaneously causally related volume. The speed of light can
exceed the conventional speed of light in non-inertial frames. The
requirement that signals can propagate results in an upper limit
on the four force squared $f^2$, which describes non-inertial
transformations. Since there is no lower bound on $f^2$, there is
in principle no upper limit on the group velocity, such that
superluminal propagation is allowed within our theory. When the
non-inertial frame of a test particle is put `on-shell', such that
the four momentum-energy squared is given by the particle's mass,
$p^2 = -m^2c^2$, then $r_{\rm max}\rightarrow G_Nm/c^2$ becomes
one half of the Schwarzschild radius. Our analysis is based on the
geodesic equation which does not take account of the self-gravity
of test particles. This suggests that the above mentioned
violation of causality will get hidden within the corresponding
particle's black hole radius, possibly rendering any violation of
causality unobservable. In conclusion, only a more proper study of
this phenomenon can fully resolve the question of causality in
Hermitian gravity.

  \item[{\tt 5.}] We define an action principle for gravity
 and matter, where we describe the matter by two scalar fields.
 The pure gravity action is holomorphic in the sense that the
 tetrad field is a holomorphic function. The reciprocity symmetry
 is imposed by a constraint action, such that it is realized
 at the level of the equations of motion (on-shell).
 This assures that the Bianchi identities are satisfied.
The scalar field action is covariant and built out of scalar
fields that are holomorphic functions (of $z^\mu$). One scalar
field has a Hermitian kinetic term, and another a holomorphic
kinetic term; the potential is the product of a holomorphic
function and its anti-holomorphic counterpart. Both scalars obey
the covariant stress-energy conservation law, such that the
Hermitian Einstein equations with scalar matter are consistent.

 \item[{\tt 6.}] We study the general relativistic limit of the theory,
 which is realized by projecting the dynamics onto the four dimensional
 space-time hypersurface. An essential element in this projection are the
 Cauchy-Riemann equations, which are a consequence of the reciprocity
 (holomorphy) symmetry of the theory. The resulting projected theory is
 {\it holographic} in the sense that, having a complete knowledge
 of the (complex) tetrad projected onto the four dimensional
 space-time manifold, allows for an unambiguous reconstruction of the full
 eight dimensional dynamics of Hermitian gravity
 (the reconstruction is essentially based on the principle of
 analytic extension generalized to Hermitian manifolds).
 We find that -- to leading order in momentum-energy $p^\mu$ --
 the geodesic equations
 reduce to those of general relativity. On the other hand,
 the (projected) dynamical (Einstein's) equations
 are not mutually identical even at zeroth
 order in $p^\mu$; the Cauchy-Riemann
 equations mix different orders of $p^\mu$.
 Thus in order to check the validity of our
 Hermitian formulation of gravity, one ought to explicitly
construct and study the Hermitian analogues of {\it each} of the
important
 solutions of general relativity. Only such a detailed comparison can
 establish the validity of Hermitian gravity, or rule it out.

 \item[{\tt 7.}] In order to investigate whether our Hermitian gravity
is a viable alternative to general relativity, we study some
important aspects of Hermitian cosmology. For definiteness and
simplicity, we focus on flat, homogeneous and isotropic universes
which expand according the power law. This class of solutions
includes most of the important cosmological solutions, including
the matter era, radiation era, inflation, and -- as a limit -- de
Sitter space. As said before, our matter is described by two
scalar fields. The purpose of the scalar field with a Hermitian
kinetic term is to satisfy the constraints of the Hermitian sector
of the theory. This field is used to `mark' the scale factor of
the Universe. The scalar field with a holomorphic kinetic term
drives the Universe's expansion. We show that at late times, when
$t\gg (G_N/c^4)E$, Hermitian cosmology reduces to FLRW cosmology
of general relativity, where $E$ denotes the relevant energy
scale. At early times the two theories deviate significantly.
While Einstein's theory exhibits the well known Big Bang
singularity, where the curvature invariants diverge, and the
theory stops giving reliable predictions, our Hermitian gravity
predicts a {\it bounce} Universe with a calculable minimal size
and maximal space-time curvature. The contracting and expanding
phases of an Hermitian gravity bounce can be asymmetric. This is a
consequence of time reversal violation induced if the scalar field
that drives the expansion violates CP (CPT is conserved). There is
a caveat though: the observers which do not exhibit a mixing
between the time-like and energy-like coordinates might still
experience a Big Bang singularity. However, such observers are
rare, and represent a negligible class of observers with very
special initial conditions (mathematically speaking, the phase
space corresponding to these observers is of measure zero).
Moreover the time-energy rotation can be absent only in those
universes where the mixing between time and energy is not
dynamically generated. Yet there is no reason to presume that our
Universe does not contain such a dynamical mixing.

 \item[{\tt 8.}]
Our analysis of Hermitian cosmology confirms the expectation that,
even at zeroth order, Hermitian gravity differs from Einstein's
gravity. The difference becomes significant, however, when
space-time curvature is large, which is still in essence an
untested sector of Einstein's theory. In future work we hope to
investigate other aspects of the theory, whenever space-time
curvature is large, such that the difference between the two
theories can again become significant, e.g. various types of black
hole solutions.

 \item[{\tt 9.}] We consider the cosmological constant
problem within our theory: the pure Hermitian gravity and two
holomorphic scalar fields in a cosmological setting. Our analysis
shows that any cosmological constant is forbidden at the classical
level, thus solving the gravitational hierarchy problem within
this framework. While this is a very welcome property of the
theory, it is still to a large extent a mystery, and awaits a
further and deeper understanding. In particular, we are interested
in the question whether a link can be established between the
reciprocity symmetry and the vanishing of cosmological constant.
Moreover, we would like to find out whether the cosmological
constant vanishes when other kinds of matter fields (in particular
fermionic and gauge fields) are included. Furthermore, we would
like to investigate whether our proof can be extended to include
quantum effects.

 \item[{\tt 10.}] Finally, we are of course
interested in quantizing Hermitian gravity. At this stage we
stress the curious fact that the commutation relations -- when
imposed on the space-time and momentum-energy
coordinates~(\ref{eq:commutation relations}) -- respect the
reciprocity symmetry. This is an important hint on how to quantize
Hermitian gravity.

 \end{itemize}

\noindent There are various other open questions which we have not
addressed here. They include: (1) can violation of the principles
of equivalence and covariance be observed; (2) can Hermitian
gravity describe the observed inwards spiralling of the
Taylor-Hulse binary pulsar; (3) does our theory meet all of the
Solar system tests; (4) is the bending of light consistent with
the predicted bending by general relativity; (5) can Hermitian
cosmology produce cosmological perturbations consistent with
observations, {\it etc.}

\section*{Acknowledgements}
\noindent We have enormously benefited from discussions with
Willem Westra, who was actively participating at the early stages
of this
project. 
The authors acknowledge financial support by FOM grant 07PR2522
and by Utrecht University.

\bibliographystyle{apsrev}

\begin{thebibliography}{99}
\bibliographystyle{unsrt}

\bibitem{Mao:2006bb}
  Y.~Mao, M.~Tegmark, A.~Guth and S.~Cabi,
  ``Constraining Torsion with Gravity Probe B,''
  Phys.\ Rev.\  D {\bf 76}, 104029 (2007)
  [arXiv:gr-qc/0608121].

\bibitem{Moffat:1978tr}
  J.~W.~Moffat,
  ``New Theory Of Gravitation,''
  Phys.\ Rev.\  D {\bf 19} (1979) 3554.

\bibitem{Janssen:2006jx}
  T.~Janssen and T.~Prokopec,
  ``Problems and hopes in nonsymmetric gravity,''
  J.\ Phys.\ A  {\bf 40} (2007) 7067
  [arXiv:gr-qc/0611005].

  T.~Janssen and T.~Prokopec,
  ``Instabilities in the nonsymmetric theory of gravitation,''
  Class.\ Quant.\ Grav.\  {\bf 23} (2006) 4967
  [arXiv:gr-qc/0604094].

\bibitem{Damour:1991ru}
  T.~Damour, S.~Deser and J.~G.~McCarthy,
  ``Theoretical problems in nonsymmetric gravitational theory,''
  Phys.\ Rev.\  D {\bf 45} (1992) 3289.

  T.~Damour, S.~Deser and J.~G.~McCarthy,
  ``Nonsymmetric Gravity Theories: Inconsistencies And A Cure,''
  Phys.\ Rev.\  D {\bf 47} (1993) 1541
  [arXiv:gr-qc/9207003].

\bibitem{Born1938Reciprocity}
  M.~Born,
  ``A Suggestion For Unifying Quantum Theory and Relativity,''
  Royal Society of London Proceedings Series A {\bf 165} (1938).

  M.~Born,
  ``Reciprocity Theory of Elementary Particles,''
  Reviews of Modern Physics {\bf 21} (1949) 463-473.

\bibitem{Einstein1905SR}
  A.~Einstein,
  ``On The Electrodynamics Of Moving Bodies,''
  Annalen der Physik {\bf XVII} (1905) 891-921.

\bibitem{Nakahara:2003nw}
  M.~Nakahara,
  ``Geometry, topology and physics,''
{\it  Boca Raton, USA: Taylor \& Francis (2003) 573 p}

\bibitem{'t Hooft:1993gx}
  G.~'t Hooft,
  ``Dimensional reduction in quantum gravity,''
  arXiv:gr-qc/9310026.

  R.~Bousso,
  ``The holographic principle,''
  Rev.\ Mod.\ Phys.\  {\bf 74} (2002) 825
  [arXiv:hep-th/0203101].

\bibitem{Einstein:1945eu}
  A.~Einstein,
  ``A generalization of the relativistic theory of gravitation,''
  Annals Math.\  {\bf 46} (1945) 578.

  A.~Einstein and E.~G.~Strauss,
  ``A generalization of the relativistic theory of gravitation. 2,''
  Annals Math.\  {\bf 47} (1946) 731.

  A.~Einstein,
  ``A Generalized Theory of Gravitation,''
  Rev.\ Mod.\ Phys.\  {\bf 20} (1948) 35.


\bibitem{Low:2005tb}
  S.~G.~Low,
  ``Reciprocal relativity of noninertial frames and the quaplectic group,''
  Found.\ Phys.\  {\bf 36}, 1036 (2006)
  [arXiv:math-ph/0506031].

   S.~G.~Low,
   ``Reciprocal relativity of noninertial frames: quantum mechanics,''
   J.~Phys.~A: Math. Theor. {\bf 40} (2007) 3999-4016
   [arXiv:math-ph/0606015].

   S.~G.~Low,
  ``Relativity group for noninertial frames in Hamilton's mechanics,''
  J. Math. Phys. {\bf 48} (2007) 102901
  [arXiv:0705.2030]

  J.~Govaerts, P.~D.~Jarvis, S.~O.~Morgan, S.~G.L~ow,
  ``World-line Quantisation of a Reciprocally Invariant System,''
  J. Phys. A: Math.~Theor.~{\bf 40} (2007) 12095-12111
  [arXiv:0706.3736].

\bibitem{Moffat:1992ud}
  J.~W.~Moffat,
  ``Superluminary Universe: A Possible Solution To The Initial Value Problem In
  Cosmology,''
  Int.\ J.\ Mod.\ Phys.\  D {\bf 2} (1993) 351
  [arXiv:gr-qc/9211020].

\bibitem{Barrow:1999jq}
  J.~D.~Barrow and J.~Magueijo,
  ``Solving the flatness and quasi-flatness problems in Brans-Dicke
  cosmologies with a varying light speed,''
  Class.\ Quant.\ Grav.\  {\bf 16} (1999) 1435
  [arXiv:astro-ph/9901049].


\bibitem{Chamseddine:2005at}
  A.~H.~Chamseddine,
  ``Hermitian geometry and complex space-time,''
  Commun.\ Math.\ Phys.\  {\bf 264} (2006) 291
  [arXiv:hep-th/0503048].

  A.~H.~Chamseddine,
  ``Gravity in complex Hermitian space-time,''
  arXiv:hep-th/0610099.


\bibitem{MantzThesis:2007}
   C.~L.~M.~Mantz,
  ``Holomorphic Gravity,''
  Utrecht University Master's Thesis (2007)
  {\tt http://www1.phys.uu.nl/ wwwitf/Teaching/2007/Mantz.pdf.}

\bibitem{Prokopec:2006kr}
  T.~Prokopec and W.~Valkenburg,
  ``Antisymmetric metric field as dark matter,''
  arXiv:astro-ph/0606315.

  T.~Prokopec and W.~Valkenburg,
  ``The cosmology of the nonsymmetric theory of gravitation,''
  Phys.\ Lett.\  B {\bf 636} (2006) 1
  [arXiv:astro-ph/0503289].


\end{thebibliography}

%
%

\end{document}